\documentclass{JHEP3} 

\usepackage{epsfig,multicol,bbm}
\usepackage{amsmath,feynmp}
\usepackage{url}
\newcommand\fverb{\setbox\pippobox=\hbox\bgroup\verb}
\newcommand\fverbdo{\egroup\medskip\noindent%
                        \fbox{\unhbox\pippobox}\ }
\newcommand\fverbit{\egroup\item[\fbox{\unhbox\pippobox}]}
\newbox\pippobox
\newcommand{\Analysis}{{\tt ExRootAnalysis}}

\newcommand{\etc}{{\it etc.}}
\newcommand{\ie}{{\it i.e.}}
\newcommand{\eg}{{\it e.g.}}

\newcommand{\ROOT}{{ROOT}}
\newcommand{\tree}{{\ROOT} tree}

\title{MadGraph/MadEvent v4: The new web generation}

\author{Johan Alwall \\
Stanford Linear Accelerator Center, Stanford University, \\
Stanford, CA 94309, USA\\
E-mail: \email{alwall@slac.stanford.edu}
}

\author{
Pavel Demin,          
Simon de Visscher,
Rikkert Frederix, 
Michel Herquet,
Fabio Maltoni\\ 
Centre for Particle Physics and Phenomenology (CP3),\\
Universit\'{e} Catholique de Louvain, \\
Chemin du Cyclotron 2, 1348 Louvain-la-Neuve, Belgium\\
E-mails: \email{simon.de.visscher@fynu.ucl.ac.be},
\email{pavel.demin@fynu.ucl.ac.be},
\email{frederix@fyma.ucl.ac.be}, 
\email{mherquet@fyma.ucl.ac.be}, 
\email{maltoni@fyma.ucl.ac.be}
}

\author{Tilman Plehn\\
School of Physics, University of Edinburgh, \\
James Clerk Maxwell Building, The King's Buildings, Edinburgh EH9 3JZ, Scotland, UK\\
E-mail: \email{tilman.plehn@cern.ch}
}

\author{David L. Rainwater\\
Department of Physics and Astronomy, University of Rochester, \\
Rochester, NY 14627, USA\\
E-mail: \email{rain@pas.rochester.edu}
}

\author{Tim Stelzer\\
Department of Physics, University of Illinois at Urbana-Champaign, \\ 
1110 West Green Street, Urbana, IL 61801, USA\\
E-mail: \email{tstelzer@uiuc.edu}
}

\preprint{CP3-07-17}      

\abstract{
We present the latest developments of the MadGraph/MadEvent Monte
Carlo event generator and several applications to hadron collider
physics.  In the current version events at the parton, hadron and
detector level can be generated directly from a web interface, for
arbitrary processes in the Standard Model and in several physics
scenarios beyond it (HEFT, MSSM, 2HDM). The most important additions
are: a new framework for implementing user-defined new physics models;
a standalone running mode for creating and testing matrix elements;
generation of events corresponding to different processes, such as
signal(s) and backgrounds, in the same run; two platforms for data
analysis, where events are accessible at the parton, hadron and
detector level; and the generation of inclusive multi-jet samples by
combining parton-level events with parton showers.  To illustrate the
new capabilities of the package some applications to hadron collider
physics are presented:
\begin{itemize}
\item[I.]  Higgs search in $pp \to H \to W^+W^-$: signal and backgrounds.\\[-23pt]
\item[II.] Higgs CP properties: $pp \to H jj$ in the HEFT.\\[-23pt]
\item[III.] Spin of a new resonance from lepton angular distributions.\\[-23pt]
\item[IV.] Single-top and Higgs associated production in a generic 2HDM.\\[-23pt]
\item[V.] Comparison of strong SUSY pair production at the SPS points.\\[-23pt]
\item[VI.] Inclusive $W+$jets matched samples: comparison with the Tevatron data.\\[-23pt]
\end{itemize}
}

\begin{document} 

\newpage
\section{Introduction}

Accurate simulation of both signal and background will play a
key role in making discoveries at the LHC. A well-known example is
given by supersymmetric models, one of the most studied and cleanest templates for
physics beyond the Standard Model. Characteristic signatures, from
large rate multi-jet plus missing $E_T$ events coming from squark pair
production, to extra $b$-jets and $\tau's$ from Higgs production, have
in general large Standard Model backgrounds that need to be measured
from the data and/or well described by Monte Carlo's. At the same
time, the most important distinctive features of the signal, such as,
for instance, mass distributions, kinematic edges and angular
correlations will be exploited not only to improve the signal over
background ratio but also to identify the quantum numbers (\eg, spin,
color) of the intermediate heavy states which decay.

The need for better simulation tools has spurred an intense activity
over the last five years, that has resulted in several important
advances in our ability to accurately simulate hard interactions. At
the matrix element level, these include the development of general
purpose event generators, such as MadGraph~\cite{Stelzer:1994ta} 
and MadEvent~\cite{Maltoni:2002qb}, CompHEP/CalcHEP~\cite{Boos:2004kh,Pukhov:2004ca},
SHERPA~\cite{Gleisberg:2003xi} and WHIZARD~\cite{Kilian:2001qz}, high
efficiency multiparton generators which go beyond the usual Feynman
diagram techniques, such as ALPGEN~\cite{Mangano:2002ea} and
HELAC~\cite{Papadopoulos:2006mh}, as well as Monte Carlo's that
include NLO corrections, such as MCFM~\cite{Campbell:1999ah} and
MC@NLO~\cite{Frixione:2002ik}.

An accurate simulation of a hadronic collision requires a careful
integration of the matrix element hard process with the full parton
showering and hadronization infrastructure \cite{Sjostrand:2006za, Corcella:2000bw}. Here
too significant advances have been made in the development of matching
algorithms such that by Catani,Krauss, Kuhn and Webber
(CKKW)~\cite{Catani:2001cc,Krauss:2002up,Mrenna:2003if}, by Mangano
(MLM)~\cite{Mangano:2006rw} and by Lavesson and Lonnblad~\cite{Lavesson:2005xu} and in
their comparison~\cite{Hoche:2006ph,matchcomp}.  
A breakthrough was also achieved by Frixione, Webber and
Nason~\cite{Frixione:2002ik,Frixione:2003ei} who showed how to
correctly interface an NLO calculation with a parton shower to avoid
double counting, releasing the first event generator at NLO,
MC@NLO. Further developments and refinements in these directions are
ongoing~\cite{Nason:2004rx,Nason:2006hf}.

While automatization has not yet been achieved at NLO, at tree-level 
we are now in a position where the matrix elements can be
calculated by several different tools, and integrated in multiple ways
into different parton-shower/hadronization codes to provide multiple
checks of the accuracy of the simulation. 

Each of the tools mentioned above was developed with a unique approach
optimized to the meet the authors intentions. This diversity of
approaches, that include a wide range of overlap in physics reach is a
key element to strengthening the programs and providing confidence in
their results. It is important to understand the philosophy/intentions
of the authors to understand and utilize their code.

The single underlying principle in the development of MadGraph and
MadEvent was to develop a tool that would maximize the amount of time
the physicist could concentrate on physics, and minimize the obstacles
between having an inspired idea, and being able to compare it to
experimental data. This originated first with MadGraph and then with
MadEvent.  At a time when many phenomenologists were spending enormous
amounts of time and energy performing important tree level
calculations, MadGraph/MadEvent was able to automate the whole
process, from the calculation of matrix elements to the generation of
unweighted parton-level events, allowing physicists to concentrate on
other pressing issues. 

This current release of the MadGraph/MadEvent package is the natural 
extension of this project both in the theoretical, and the experimental
directions. While the original MadGraph had the standard model
``hard-coded'' this version includes several new models (MSSM, 2HDM,
HEFT,\ldots), as well as the capability for user defined models. It
also includes the ability to seamlessly pass the events through a full
hadronic simulation so they can be subjected to a complete detector
simulation. Utilizing the strengths of the web, a physicists can now
literally go from an inspired concept to a full event simulation with
detector reconstruction with just a few clicks of the mouse. In
addition, in order to facilitate the communication between theorists
and experimentalists, the complete process from diagram generation to
event analysis can be recreated from a simple text file which collects
all the input cards used in the various phases of the simulation.

In this work we briefly illustrate the new features of the code,
focusing more on the structure and on general aspects than on the
technical details. In Sections 2 and 3 we give an overview of the
package. In Section 4 we discuss the models for which a dedicated
implementation exists (SM+HEFT, MSSM, 2HDM) as well as a framework
where new models can be implemented starting from the SM. In Section 6
our approach to generate multi-jet samples through matching is
described. Section 7 collects the descriptions of some of the tools
available that allow the users to handle the parton-level events in
the Les Houches format for different purposes: from analysis to
further simulation, to the decay of unstable particles maintaining the
leading spin-correlations. In Section 8 we provide several examples,
from SM measurements to the search for SUSY in inclusive signatures,
of studies that can be performed with our package. We draw our
conclusions and present our current line of research and development
in the final section.

\section{The path to event generation}
\label{sec:event_generation}

The new structure of the MadGraph/MadEvent package is shown in
Fig.~\ref{fig:chart}. Each of the four steps (code creation,
parton-level event, hadron-level event and reconstructed object
generation) are driven by input cards provided by the user. All cards have
a comment section which describes in detail their content and
syntax. Here we just mention their functionalities and how the
generation proceeds, as outlined in Fig.~\ref{fig:chart}.  For sake of
simplicity, we consider the procedure to be followed by a web user,
who can control the full generation process on one of the
MadGraph clusters, via a web interface. The steps undertaken by a user
working locally on his/her own computer, are the same, but in this
case the simulation is initiated by calling scripts from the command
line. In this case, the 
exact syntax to be used can be found in the corresponding {\tt README}
files and in our on-line documentation.

\FIGURE[ht]{
\epsfig{file=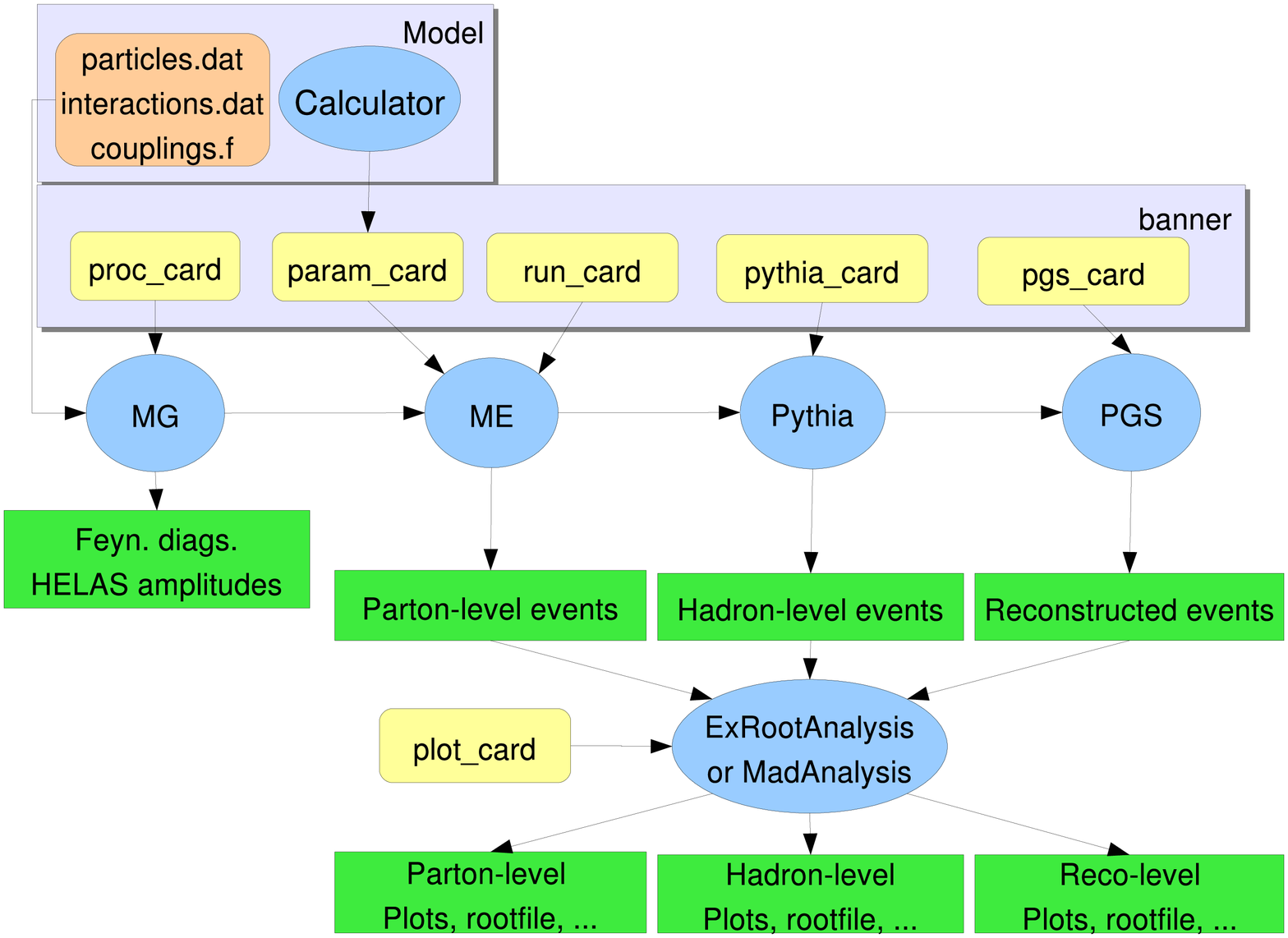, width=15cm}
\caption{Flow chart of the MadGraph/MadEvent event generation.
Each step is controlled by a specific card. Details are given in the text.}
\label{fig:chart}
}

The first phase is the creation of the code specific to the 
process that the user requests. This can be done directly by filling a
web form or 
by editing the process card, {\tt proc\_card.dat}, to identify 
the process or processes to be included in the code, together
with the model and the maximum order of the couplings to appear in the 
tree-level diagrams.  Labels can be specified to identify multi-particles
(such as jets) which defines summations over physical
particles. Examples of process cards are given in the
Section~\ref{sec:applications}.  Once 
completed, the process card is read by MadGraph and a process-specific code
(packaged in a gzipped tar file) is created and can either be downloaded
in order to run the simulation locally, or be run directly on the
cluster using a web interface.

Two cards are needed to perform the event generation.  The first one is
the parameter card, {\tt param\_card.dat}, which contain the numerical values of the
necessary parameters for a specific model.  The
parameter card has a format compliant with the SUSY Les Houches Accord
and it is dependent on the physics model. At variance with the
previous version, secondary parameters and widths are no longer
computed in MadEvent, but instead passed through the parameter card
also for the Standard Model.  Such a card can be generated with an
independent code, which we refer to as a ``Model Calculator''. There
are calculators available on the web for the SM, MSSM and 2HDM models,
which, starting from the parameters in the Lagrangian (primary
parameters) calculate all secondary parameters (such as masses,
widths, and auxiliary parameters) needed by MadGraph to perform the
cross section integration. We stress that the calculation of all such
parameters should be done by a Model Calculator since they must be
consistently performed at tree level.  The second card to be completed
is the run card, {\tt run\_card.dat}, where all the information
regarding the event generation (number of events requested, random
seed,\ldots), the collision (energy, beam type, pdf's, scales) and the
acceptance of the detector ($p_T^{min}$, $\eta_{\max}$, \ldots) is
passed. Both cards can be generated using a web form, or
uploaded. Once the cards are completed, the event generation can
start.

As a result of the run, the user will obtain a data file with the 
unweighed events in the Les Houches format, a set of plots of
kinematical variables, and a ROOT file.

The parton-level events can then be passed to Pythia by specifying in
the {\tt pythia\_card.dat} the desired options. Events at the hadron
level are then passed to PGS, whose parameters are set in the {\tt
pgs\_card.dat}. On the web the Pythia and PGS cards can be completed
at the same time as the parameter and run cards or at a later time. In
the first case the generation goes all the way through the
reconstructed objects at the detector level, and the information
(events and plots) of intermediate stages of the simulation is kept.

The  MadGraph/MadEvent package can be used at different levels, depending on
the user's needs/expertise. We have identified three main categories:
\begin{itemize}

\item A {\bf web user} performs the full generation
(from the process specification to the Pythia or PGS events) over the
Web, downloading only the input cards and the results (events and
plots). No code download or installation is needed as all phases are
handled via a web interface and the generation is performed on one
of our clusters. Each user manages a personal database where all the
codes/runs/events are stored for future use.  This is the simplest
level of accessing MadGraph/MadEvent, and it is suitable to both
theorists and experimentalists who might want a quick determination of
a cross section or a fair amount of events to conduct preliminary
studies using the available physics models.  Code creation and all other
tools on the web page are accessible to anybody upon web
registration.

For a web user it is enough to understand the logic and the steps of the code,
Fig.~\ref{fig:chart}, and to correctly manage the information which 
is passed through the input cards. This is explained in detail 
in our on-line documentation.

\item A {\bf local user} creates the code specific to the process of
interest over the web and then downloads and runs it locally on
her/his personal computer or cluster. This allows somewhat
more flexibility (for instance, special selection cuts can be implemented,
as well as different PDF's, dynamic scale choices, and so on), as well as
the possibility for the production of large samples of events. In this case,
it is suggested to run in parallel over a cluster with
the PBS~\cite{pbs} or CONDOR~\cite{condor} batching systems. 

\item A  {\bf developer} downloads the full MG/ME package
to exploit all functionalities locally.  This is certainly the
preferred option not only for the users making detailed and/or
sophisticated simulation studies, but also for developers of new tools
or models in the MG/ME framework. We underline in passing that
experiments or larger collaborations that have at their disposal a
computer farm running a batching system, such as PBS or CONDOR, might
also install the web server software, which manages all the generation
steps (code creation, event generation, interfacing to Pythia and PGS,
as well as ROOT-based analysis) and user databases automatically. The
corresponding web server software and tech support can be obtained
from the authors.
\end{itemize}

The installation of the MG/ME package  is straightforward since no
external libraries are needed. The interested reader should refer to the
README file contained in the main directory, obtained after
decompressing the {\tt MG\_ME\_V4.X.Y.tar.gz} archive file. The main
directory is organized in subdirectories whose contents we briefly
describe below.
\begin{itemize}
\item
{\tt MadGraphII} : The source code (Fortran) and executable ({\tt
mg2}) for the matrix-element generator.  {\tt mg2} generates not only
a code for the matrix element squared for a given process, but also
all subprocesses contributing to it. When interfaced with MadEvent it
also generates the mappings needed for the phase space integration,
including the Single-Diagram-Enhanced expressions. {\tt mg2\_sa} is
the ``stand alone'' version of MadGraph producing the code for the
matrix element squared only.
\item
{\tt Template} : The general structure of a process dependent
code, as
shown in Table~\ref{tab:template}.  To create the code corresponding
to a new process, a copy of the full directory should be made (at the
main directory level) and the script {\tt ./bin/newprocess} invoked
from it.
\item {\tt Models} :  It contains the available models (each of them
is a directory) including the user model template ({\tt usrmod}). The
user developing a new model makes a copy of the {\tt usrmod} directory
at this level, giving a name to it that identifies the model (which
can later be set in {\tt proc\_card.dat}).
\item{\tt HELAS} :  The HELAS library~\cite{Murayama:1992gi} source code.
\item{\tt DECAY} : The source code to decay unstable particles directly from 
parton-level LHE files.
\end{itemize}

\TABLE[t]{
\label{tab:template}
\begin{tabular}{|c|c|} \hline
{\tt Dir name}      & Content\\
\hline
{\tt Cards }         & steering cards\\
{\tt Source }       & proc. independent source  files \\
{\tt SubProcesses } & proc. dependent source files and dirs for the subprocesses\\
{\tt Events }        & LHE event, plots and ROOT files\\
{\tt bin }              & executables and scripts (csh and Perl)\\
{\tt lib }               & libraries and PDF data\\
{\tt HTML}          & web pages\\
\hline
\end{tabular}
\caption{Organization of the process directory (Template). When the code
specific to a user's process is generated over the web, it is packaged
into {\tt MadEvent.tar.gz} which has the structure above.}  }

\section{The Standard Model implementation}
\label{sec:sm}
\subsection{Standard Model interactions}
\label{sm}

The Standard Model of particles and interactions, based on the
$SU(3)_c \times SU(2)_L \times U(1)_Y$ gauge symmetry has been
available since the first versions of both MadGraph and more recently
of MadEvent.  There is, however, one important differences w.r.t.  the
previous version of the package, regarding how the couplings of the
models are handled.  As was already mentioned in the previous section,
the task of computing from the parameters in the Lagrangian (primary
parameters) all the secondary parameters (masses, widths and dependent
parameters) needed by MadGraph is left to an external program, the SM
Calculator. The output of the SM Calculator is a parameter card, {\tt
param\_card.dat}, which contains the numerical values of the main
couplings (primary and secondary) of a specific model. The parameter
card has a format compliant with the SUSY Les Houches Accord.

A simple example is given by the EW parameters that characterize
the gauge $SU(2)_L \times U(1)_Y$ interactions and its breaking:
in the Standard Model there are five relevant parameters, 
$\alpha_{em}, G_F, \sin \theta _W, m_Z,m_W$ of which
only three are independent at tree level. Various schemes 
differing by the choice of the parameters considered independent are
used in the literature. In the SM Calculator, the default is to take
$G_F,m_Z,m_W$ as inputs and derive $\alpha_{em}, \sin \theta_W$,
but other choices are available. As a result a consistent and unique set of 
values of the couplings appearing in the Feynman rules is derived and
used for the computation of the amplitudes.

Another sometimes important feature of our SM implementation, is the
possibility of distinguishing between the kinematic mass (pole mass)
for the quarks and that entering in the Yukawa coupling definition 
($\overline{MS}$ mass). For the latter, the user can choose to evolve
the mass to the scale corresponding to the Higgs mass, which leads to
an improvement of the perturbative expansion.

Finally, we mention that various versions of the Standard Model are
actually available for specific studies. For example, in the ``minimal
SM'' ({\tt sm}) the CKM matrix is diagonal while in the {\tt sm\_ckm}
model a mixing between the first and second generation is allowed
(Cabibbo angle).  Another example is the {\tt sm\_nohiggs} model where
the Higgs has been eliminated and the EWSB sector behaves as a
non-linear sigma-model.

\subsection{Higgs Effective Theory}
\label{heft}

The Higgs effective field theory (\verb heft ) model is an `extension'
of the Standard Model (SM), where the Higgs boson couples directly to
gluons (and photons) \cite{Shifman:1979eb,Kniehl:1995tn,Dawson:1993qf}.  
In the SM these couplings are present through a
heavy (top) quark loop. For a not too heavy Higgs ($m_h<2m_t$), it is
a good approximation to take the mass of the heavy quark in the loop
to infinity\footnote{For this approximation to hold, not only should the Higgs mass
be smaller than twice the top mass, also all other kinematic variables,
such as the transverse momentum of the Higgs boson, should be smaller than
$2m_t$ \cite{DelDuca:2001eu}.}.  This results in effective couplings between gluons and
Higgs bosons.

The effective vertices can be derived from the effective dimension five Lagrangian
\begin{equation}
  \label{higgs1}
  \mathcal{L}_{h}=-\frac{1}{4}g_hG_{\mu\nu}^aG_{\mu\nu}^a \Phi,
\end{equation}
where $G^a_{\mu\nu}=\partial_{\mu}A^a_{\nu}-\partial_{\nu}A^a_{\mu}+gf^{abc}A^b_{\mu}A^c_{\nu}$. The coupling constant $g_h$ is given by 
\begin{equation}
  \label{eq:55}
  g_h=\frac{\alpha_s}{3\pi v}.
\end{equation}
Due to the non-abelian nature of the $SU(3)_C$ color group the
effective vertices do not only include two, but also three and four
gluons coupling to the Higgs boson. Since MadGraph can work only with
three-- and four--point vertices, the four--gluon interactions in the
{\tt heft} model are obtained by rewriting the QCD four-gluon
interaction in terms of three--point vertices with an extra
\emph{non-propagating} internal tensor particle,
$T$~\cite{Caravaglios:1998yr,Pukhov:1999gg}. This trick can be easily
understood by noting that the usual (text-book) form of the four-gluon
interaction is the sum of three terms, whose color and Lorentz
structure correspond to $2 \to 2 $ diagrams where a color octet tensor
is exchanged in the $s,t,u$ channels.  With the introduction of this
extra particle, the four--gluon--Higgs vertices can be reduced to
diagrams with at most four-point vertices. To get the standard
diagrammatic visualization of four--gluon and four--gluon--Higgs
vertices it is sufficient to contract the $T$ particle lines to
a single point.

The gluon couplings to a pseudo--scalar Higgs are also implemented. The
name of the pseudo-scalar Higgs in MadGraph is {\tt h3} (\ie,
~the same as in the 2HDM and MSSM models). The effective dimension five
Lagrangian for the pseudo-scalar Higgs coupling to the gluons is
\begin{equation}
  \label{eq:85}
    \mathcal{L}_{A}=\frac{1}{2}g_AG_{\mu\nu}^a\tilde{G}_{\mu\nu}^a \Phi_A,
\end{equation}
where $\tilde{G}_{\mu\nu}^a$ is the dual of $G_{\mu\nu}^a$, $\tilde{G}_{\mu\nu}^a=\frac{1}{2}\epsilon^{\mu\nu\rho\sigma}G_{\rho\sigma}^a$. 
The effective coupling constant $g_A$ is given by
\begin{equation}
  \label{eq:93}
  g_A=\frac{\alpha_s}{2\pi v}.
\end{equation}
The pseudo--scalar Higgs has only effective couplings to two or three
gluons. The four--gluon--pseudo--scalar Higgs vertex is absent due to
the anti-symmetry of the epsilon tensor $\epsilon^{\mu\nu\rho\sigma}$.
If a mixed Higgs with no definite CP parity is needed,
it sufficient  to change the couplings of
the Higgs to the gluons. First  generate the process with
the SM Higgs, then, after downloading the code, 
change the coupling in the {\tt ./Source/Model/couplings.f}\ file. The
coupling constant is defined as a two--dimensional object, where the
first and second elements are the CP--even and CP--odd couplings
of the Higgs to the gluons, respectively. The
HELAS subroutines automatically use the correct kinematics for
odd--, even-- or mixed CP Higgs's coupling to the gluons.
At present, the implementation allows production of only one
Higgs-boson.  The effective couplings of two Higgs bosons to gluons are
available in HELAS, but not yet included in the HEFT model.

\section{Going beyond the Standard Model}

In the current version a few new physics models
have been added and fully tested: 
the minimal supersymmetric extension of the Standard
Model (MSSM) and the general two Higgs doublet model
(2HDM). In addition, a framework for setting up new models is available.
Here, we briefly describe the main features of these new implementations.

\subsection{The 2HDM implementation}
The two-Higgs-doublet model (2HDM) has been extensively studied for
more than twenty years, even though it has often been only
considered as the scalar sector of larger models like the MSSM
\cite{Gunion:1989we} or Little Higgs models
\cite{Arkani-Hamed:2002qx}. The generic 2HDM considered here may
display by itself an interesting phenomenology justifying its
study. As a non exhaustive list, let us mention new sources of $CP$
violation in scalar-scalars interactions \cite{Branco:1999fs},
tree-level flavor changing neutral currents (FCNCs) due to non
diagonal Yukawa interactions, dark matter candidates
\cite{Barbieri:2006dq} or Higgs bosons lighter than the LEP bound
\cite{Gerard:2007kn}.

In the ``full'' version of the model ({\tt 2hdm\_full}), no particular
restrictions are imposed on the interactions allowed by gauge
invariance, except electric charge conservation. Many diagrams
involving tree-level FCNCs and violating the $CP$ symmetry are thus
present. The user who is not interested in these phenomena should use
the ``simplified'' version of the model ({\tt 2hdm}), where the number
of generated diagrams is in general much smaller.

The following naming convention is used: {\tt h+} and {\tt h-} stand
for the positively and negatively charged Higgs bosons and {\tt h1},
{\tt h2} and {\tt h3} stand for the neutral ones. Since the $CP$
invariance of the potential is not assumed, the neutral bosons are not
necessarily $CP$ eigenstates and the standard naming convention in
this case (\ie, {\tt h1} being the lightest one and {\tt h3} the
heaviest one) is used.

TwoHiggsCalc is the calculator associated with the model. It has been
written in {\tt C} and is accessible from a web interface. It has been
designed to compute input values for the 2HDM extension of
MadGraph/MadEvent but it can also be used as an independent
tool. Starting from various parameters of the Lagrangian, such as the
vacuum expectation values (vevs) or the Yukawa couplings, the program
computes useful secondary physical quantities at leading order such as
the scalar mass spectrum, the mixing matrix, the total decay widths
and the branching ratios.

TwoHiggsCalc reads input and writes out results in a specific format
close to the ``SUSY Les Houches Accord 1.0" convention for SUSY
parameters \cite{Skands:2003cj}. This format can later be read by
MadEvent to perform numerical calculations for 2HDM processes. A {\tt
README} file describing this modified version of the LHA format used
as input convention is available. To ease the use of TwoHiggsCalc, a
web form has been designed to automatize the parameter card writing
process. Numerical values for the parameters (units being fixed when
needed) can be entered on this form. Some simple algebraic expressions
can also be used. The {\tt +,-,*,/} operators and the
reserved keyword {\tt PI}, \eg, {\tt PI/2+3*PI/2}, are correctly interpreted.

In the general 2HDM, one has the freedom to choose a specific basis
for entering parameters. All the possible choices are physically
equivalent (see \eg\ \cite{Davidson:2005cw} for a
discussion). TwoHiggsCalc and the 2HDM model both assume that the
parameters are given in a particular basis, called the ``Higgs basis''
where only one Higgs doublet gets a vacuum expectation value. An
independent program , Gen2HB, has been written to convert parameters
given in an arbitrary basis (where both Higgs doublets get vevs),
called ``generic'', to parameters in the Higgs basis. See
\cite{Branco:1999fs} for more information on basis invariance and on
the notation used.

The scalar potential in the Higgs basis reads 
\begin{eqnarray*}
V &=& \mu_1 H_1^\dag H_1 +\mu_2 H_2^\dag H_2-\left(\mu_3 H_1^\dag H_2+\mathrm{h.c.}\right)\\
  & & \lambda_1 \left(H_1^\dag H_1\right)^2+ \lambda_2 \left(H_2^\dag
  H_2\right)^2\\ & & + \lambda_3 \left(H_1^\dag
  H_1\right)\left(H_2^\dag H_2\right)+ \lambda_4 \left(H_1^\dag
  H_2\right)\left(H_2^\dag H_1\right)\\ & & +\left[\left( \lambda_5
  H_1^\dag H_2 + \lambda_6 H_1^\dag H_1+ \lambda_7 H_2^\dag
  H_2\right)\left(H_1^\dag H_2\right)+\mathrm{h.c.}\right]\,.
\end{eqnarray*}
All parameters in front of quartic terms and the charged Higgs mass
are input parameters, while $\mu_1$, $\mu_2$ and $\mu_3$ are fixed by
minimization constraints and by the vev extracted from the observed SM
parameters. $\lambda_1$ to $\lambda_4$ are real while $\lambda_5$ in
general is complex. However, since only the phase differences between
$\lambda_5$, $\lambda_6$, $\lambda_7$ and $\mu_3$ matter, the phase of
$\lambda_5$ can always be rotated out. It is thus considered as a real
parameter while $\lambda_6$ and $\lambda_7$ are a priori complex.

In the same basis, the Yukawa interactions read
\begin{eqnarray*}
\mathcal{L}_Y&=&\frac{\overline{Q_L}\sqrt{2}}{v}\left[(M_d H_1 +   Y_d   H_2)d_R+(M_u \tilde{H}_1 +   Y_u   \tilde{H}_2)u_R\right]\\
& &+\frac{\overline{E_L}\sqrt{2}}{v}\left[(M_e H_1 +   Y_e   H_2)e_R\right]\,.
\end{eqnarray*}
Yukawa couplings are expected to be given in the physical basis for
fermions, \ie, in the basis where the mass matrix is diagonal. Since
in the Higgs basis only the first Higgs doublet gets a non zero vev,
the $M$ matrices are completely fixed by the physical fermion masses
and CKM mixing matrix (restricted to Cabibbo angle) while the $Y$
matrices (giving the couplings of the second Higgs doublet) are a
priori free. For these matrices, the first index refers to doublet
generation while the second refers to the singlet generation. For
example, {\tt Y2B} stands for the complex Yukawa couplings of the
second Higgs doublet to the second generation quark left doublet and
to the bottom singlet.

In the generic basis, similar expressions are assumed. For the scalar
potential all parameters in front of quartic terms are inputs as well
as $\tan(\beta)$, the norm of $\mu_3$ and the phase of $v_2$.  The
overall vev is again extracted from SM parameters while mass terms
parameters, like $\mu_1$, $\mu_2$ and the phase of $\mu_3$, are fixed
by the minimization constraints. $\lambda_1$ to $\lambda_4$ are real
parameters, $\lambda_5$, $\lambda_6$ and $\lambda_7$ are a priori
complex. Like in the Higgs basis, the Yukawa couplings must be given
in the physical basis for fermions. Since the mass matrices are fixed,
only the Yukawa coupling matrices of the second Higgs doublet
($\Gamma$), is required. The other one is going to be automatically
evaluated to match observed fermion masses and CKM mixing matrix
(restricted to Cabibbo angle). For the $\Gamma$ matrix, the first
index refers to doublet generation while the second one refer to the
singlet generation.  For example, {\tt G2B} stands for the complex
Yukawa couplings of the second Higgs doublet to the second generation
quark left doublet and to the bottom singlet.

Given the above parameters and some SM parameters, TwoHiggsCalc
computes the following quantities
\begin{itemize}
\item Scalar particles mass spectrum
\item Normalized mixing matrix of neutral scalars (called $T$ in \cite{Branco:1999fs})
\item Decay widths for all scalars as well as for $W$ and $Z$ bosons
      and the top quark. All widths are evaluated at tree-level using
      the same couplings as in MadEvent. Below threshold formulas
      are included for the scalar decays into two vector bosons and
      the one loop driven scalar decay into two gluons is also
      computed.
\end{itemize}

The LHA blocks and parameters used by MadEvent are given in
Table~\ref{tab:blocks2HDM}. All blocks in the table are provided by
TwoHiggsCalc. Note that if parton density functions (PDFs) are used in
the MadEvent run, the value for $\alpha_s$ at $M_Z$ and the order of
its running is given by the PDF. Otherwise $\alpha_s(M_Z)$ is given by
block {\tt SMINPUTS}, parameter 3, and the order of running is taken
to be 2-loop. The scale where $\alpha_s$ is evaluated can be fixed or
evaluated on an event-by-event basis like in the SM.

\begin{table}
\begin{tabular}{|c|l|}
\hline
        Block & Comment \\
\hline
        {\tt SMINPUTS} & From 1 to 4, SM parameters, see the SM
        section for more details\\ 
        {\tt MGSMPARAM} & Extra block with
        $\sin\theta_W$ and $M_W$, see the SM section for more
        details\\ 
        {\tt MGYUKAWA} & ``Yukawa'' masses used in the
        Yukawa couplings evaluation\\ 
        {\tt MGCKM} & The full CKM
        matrix\\ 
        {\tt BASIS} & Basis choice, must be 1 (Higgs basis)
        for {\tt MadEvent} !\\ 
        {\tt MINPAR} & Scalar potential
        parameters in the Higgs basis\\ 
        {\tt YUKAWA2} & Yukawa
        couplings of the second Higgs doublet \\ 
        {\tt MASS} & All SM
        particles masses, plus the five new Higgs boson masses\\ 
        {\tt TMIX} & The scalar mixing matrix\\ 
        {\tt DECAY} & For all the
        Higgs bosons, top, $W^\pm$ and $Z$\\
\hline
\end{tabular}
\caption{\label{tab:blocks2HDM}LHA blocks used in the 2HDM implemention}
\end{table}

\subsection{The MSSM implementation}


One of the most popular extensions of the Standard Model is TeV scale
supersymmetry. Supersymmetry solves the problem of quadratically
divergent corrections to the Higgs boson mass by the introduction of
new bosonic particles having the same couplings as the Standard Model
fermions, and new fermions having the same couplings as the Standard
Model bosons, thus cancelling the loop contributions to the Higgs mass
to all orders. The Minimal Supersymmetric Standard Model, MSSM,
represents the minimal particle content for a supersymmetric extension
of the Standard Model together with the maximum coupling space allowed
by so-called ``soft supersymmetry breaking terms'' in the effective
low-energy Lagrangean. These are constructed not to introduce new
divergencies in any couplings, and therefore maintain the
cancellations of quadratically divergent corrections to the Higgs
mass. For an introduction to supersymmetry and the MSSM, see \eg,
Refs.~\cite{Aitchison:2005cf,Martin:1997ns}.


The implementation of the MSSM particles and vertices into MadGraph II
was made in Ref. \cite{Cho:2006sx,Hagiwara:2005wg}, following the conventions
of Refs. \cite{Gunion:1984yn} and \cite{Plehn:1998nh}. Specifically, it is
restricted to the minimal supersymmetric model conserving $R$-parity,
without {\em CP}-violating phases and with diagonal CKM and MNS
matrices. Higgs Yukawa couplings as well as mixing between right- and
left-handed sfermions are implemented only for the third
generation. However, no specific supersymmetry breaking scheme is
assumed, so the spectrum and couplings of the supersymmetric particles
can be produced with any spectrum generator regardless of the
assumptions going into its calculations. The spectrum and couplings of
the particles are read through SUSY Les Houches Accord files
\cite{Skands:2003cj}.


In order to consistently calculate decay widths and the dependent
parameters, a model calculator (see
Section~\ref{sec:event_generation}) for the MSSM is available.
MSSMCalc takes a SUSY Les Houches Accord (SLHA) file
\cite{Skands:2003cj} from any Spectrum generator as input, and
produces a MadEvent readable file, {\tt param\_card.dat}, with the
missing Standard Model parameters, as well as decay widths for all
supersymmetric particles (calculated at leading order by Sdecay
\cite{Muhlleitner:2003vg}), the Higgs particles and the top, $W^\pm$
and $Z$ particles. Care has been taken to ensure that the parameters
used in the calculation of decay widths are as similar as possible to
the parameters used in MadEvent, since the correct total decay widths are
vital to get the correct tree-level cross-sections for processes
involving decaying particles. 

In the default run mode, MSSMCalc uses the Standard Model parameters
given in the SUSY Les Houches accord ($\alpha_{em}$, $G_F$ and $M_Z$)
to calculate the parameters $\sin\theta_W$ and $M_W$, which are stored
in a MadEvent specific block {\tt MGSMPARAM} in the resulting {\tt
param\_card.dat}. The $b$ quark pole mass is calculated from the
$\overline{MS}$ mass at 2-loop order. Another option is to
extract the Standard Model parameters (and the vacuum expectation
value ratio $\tan\beta$) from the chargino and neutralino mixing
matrices, in order to ensure unitarity of ino-ino scattering at high
energy. In this mode, also the Yukawa masses of the third generation
fermions are extracted from the third generation sfermion mixing
matrices. For a thorough discussion of this option, see
section II C of \cite{Cho:2006sx}.

The strong coupling $\alpha_s$ is calculated in MSSMCalc using 2-loop
renormalisation group running in the $\overline{MS}$ scheme, at
the scale specified in the {\tt GAUGE} block statement. The value used
for the strong coupling $g$ in the decay width calculations is stored
for comparison in the block {\tt GAUGE}, parameter 3. Note however,
that the value of $\alpha_s$ used in MadEvent is given by the choice
of parton distribution function and the scale chosen in the run.

If there are blocks missing in the SLHA file which are necessary for
Mad\-Event, MSSMCalc will produce a {\tt param\_card.dat} file
containing error messages.

The SUSY Les Houches blocks and parameters used by MadEvent are given
in Table~\ref{tab:blocks}. All blocks in the table should be provided
by the user (and are indeed provided by most MSSM spectrum
generators), except for the {\tt MGSMPARAM} and the {\tt DECAY} blocks
which are produced by the parameter calculator MSSMCalc. Note that if
parton density functions (PDFs) are used in the MadEvent run, the
value for $\alpha_s$ at $M_Z$ and the order of its running is given by
the PDF. Otherwise $\alpha_s(M_Z)$ is given by block {\tt SMINPUTS},
parameter 3, and the order of running is taken to be 2-loop. The scale
where $\alpha_s$ is evaluated is however always given by the ``scale''
parameter in the {\tt run\_card.dat}.

\begin{table}
\begin{tabular}{|c|l|}
\hline
        Block           & Comment       \\
\hline
        {\tt SMINPUTS}  & Except for 5, the $b$ quark $\overline {MS}$ mass\\
        {\tt MGSMPARAM} & Extra block with $\sin\theta_W$ and $M_W$, written by MSSMCalc \\
        {\tt MASS}      & Including 5, the $b$ quark pole mass\\
        {\tt NMIX}, {\tt UMIX}, {\tt VMIX}      &       \\
        {\tt STOPMIX}, {\tt SBOTMIX}, {\tt STAUMIX} &   \\
        {\tt ALPHA}     &               \\
        {\tt HMIX} & Only parameters 1 ($\mu$) and 2 ($\tan\beta$)\\
        {\tt AU}, {\tt AD}, {\tt AE} & Only the third generation parameter 3 3\\
        {\tt YU}, {\tt YD}, {\tt YE} & Only the third generation parameter 3 3\\
        {\tt DECAY} &   For all SUSY particles, Higgs bosons, top, $W^\pm$ and $Z$\\
\hline
\end{tabular}
\caption{\label{tab:blocks}SHLA blocks used by SUSY MadEvent. See the
text for details.}
\end{table}

\subsection{The User Model}

Beside the models already implemented in the MadGraph structure, it is
also important to let the possibility to the non-expert to
be able to test his or her own models. For example, adding a $Z^{'}$
boson or a $t^{'}$ fermion to Standard Model and all their possible
couplings to other particles should be now straightforward. For this purpose a new
framework has been developed to provide users with an easy and safe, yet
flexible, tool for implementing their own models.  It is important to
mention that the user model is limited by the MadGraph and HELAS
assumptions for the  Lorentz and color structure of the vertices.

The aim here is not to give a detailed step-by-step description on how
to implement a new model (which is available in the README file in the
{\tt usrmod} directory), but to present the philosophy adopted in the
design of the framework.  In practice, a model is defined by a list of
particles with given quantum numbers and their interactions.  The
implementation of a model proceeds in three simple steps.

First, three files have to be provided: i) the list of particles; ii)
the list of the interactions (the complete Standard Model, which is
already present by default, plus the user's new inputs) iii) the list
of possible new parameters present in non-SM couplings expressions.
As a second step a Perl script, {\tt ConversionScript.pl}, is invoked
that takes the above information and creates all the files needed
by MadGraph/MadEvent. As a final step, the user must modify by hand the {\tt
couplings.f} file to set the values of the couplings and
provide the formulas to go from the parameters of the model to
those appearing in the Feynman rules.

Let us consider the implementation of a $Z'$ vector boson as an example. 
In the particle list file ({\tt particle.dat}), an extra $Z$ appears as
\begin{verbatim}
#Name anti_Name  Spin    Linetype Mass Width Color Label Model
#xxx   xxxx      SFV      WSDC     str  str  STO    str  PDG code
...
#MODEL EXTENSION
zp       zp       V        W      ZPMASS ZPWID  S   ZP   32
\end{verbatim}
\noindent where the needed information is provided in a simple
syntax. In this example we have a $Z^{'}$ ({\tt zp}) of vector type
({\tt V}), with mass and width {\tt ZPMASS} and {\tt ZPWIDTH}. We made it
a color singlet by setting color to {\tt S}. The linetype {\tt W} and
the label {\tt ZP} are used in the Feynman diagrams for cosmetics
only.

As described above, in the second file ({\tt interactions.dat}) the
interactions are specified. In this file the names of the particles as
defined above have to be used. In this example the vertex
$Z'd\overline{d}$, {\it i.e.}, the vertex between the $Z'$ and two
down quarks, is shown.
\begin{verbatim}
#   USRVertex
d d ZP  GZPD  QED
\end{verbatim}
\noindent The name {\tt GZPD} is choosen arbitrarily (but paying
attention not to choose a name already in use); {\tt QED} indicates
the type of interaction.

If there are new parameters related to coupling expressions, the user
can edit a third file {\tt VariableName.dat}. For example, to include
the couplings between $Z^{'}$ and down quarks, the expression of the
coupling appearing in the Feynman rule can be written as
\begin{equation}{\label{eqn}}
 g_{Z^{'}d\overline{d}}=-\sqrt{4\pi\alpha}\bigg[C_1\frac{1}{s_{W}c_{W}}\bigg(-\frac{1}{2} + \frac{s_{W}2}{3}\bigg)\bigg]\frac{1+\gamma_{5}}{2}+C_2\frac{s_{W}}{3c_{W}}\frac{1-\gamma_{5}}{2},
\end{equation}
where $s_W$ and $c_W$ are $\sin\theta_W$ and $\cos\theta_W$, respectively. $C_1$ and $C_2$ are one for a SM like $Z'$. This can, of course, also be written as
\begin{equation}
 g_{Z^{'}d\overline{d}}=C_1'\frac{1+\gamma_{5}}{2}+C_2'\frac{1-\gamma_{5}}{2}
\end{equation}
where no link with the parameters in the SM is explicit.
The declaration of these parameters appears as a list in {\tt VariableName.dat}
\begin{verbatim}
C1  #first variable name
C2  #other variable name
\end{verbatim}

The second step is to run the {\tt ConversionScript.pl}. The script
uses the above three files as input and creates the parameter card,
{\tt param\_card.dat} and the file {\tt couplings.f}, as well as other
files needed by MadEvent. The third and
final step is to include in these files the numerical values related
to the model.  The user model parameter card is in LHA format and is
similar as the SM one, but also includes the information about the new
particles and coupling parameters.  The masses and widths of the new
particles, as well as the new coupling parameters, need to be set by
the user to their correct numerical values.  In the {\tt couplings.f}
file, expressions for the coupling strengths have to be provided. The
Perl script already takes care of the formats for the couplings in
such a way that they are in HELAS compliant. For our example, the
expression in Eq.~(\ref{eqn}) can be implemented as
\begin{verbatim}
     GZPD(1)=dcmplx(C1*( -ez*(-Half + sin2w/Three)),Zero)
     GZPD(2)=dcmplx(C2*( -ey/Three),Zero)
\end{verbatim}

The possibility of testing the couplings values and/or other
parameters like the masses and the widths is also available. The
program {\tt testprog} can be compiled and run and it prints the
values for all masses and couplings that will be used by MadGraph. A
second program {\tt couplingsvalues} writes out the names and the
corresponding values of couplings in a format that can be read by
external tools like the width and decay calculator
BRIDGE~\cite{Meade:2007js}.

\section{Matching of jet production by parton showers and matrix elements}
\label{sec:matching}

For many processes at the LHC, the major backgrounds include multijet
production, either pure QCD production or jet production in
association with weak vector bosons or top quarks. Also when generating signal
processes, such as Higgs boson production, it is often important to
understand the jet activity, \ie\ the probability for extra jet
production, to be able to discriminate from the backgrounds. 

Parton showering, \eg\ by Pythia, is well known to give a good
description of parton emissions when the emitted partons are close in
phase space. To describe hard and well separated partons, matrix
element generators such as MadGraph/MadEvent should be used. However,
even in that case, it is necessary to perform parton showering and
hadronization in order to get a realistic description of the event, in
particular when detector simulation is needed. In order to combine
these two descriptions, it is essential to use some kind of matching
between them to avoid double counting of emissions in overlapping
phase space regions.

One such matching scheme is the CKKW algorithm
\cite{Catani:2001cc,Krauss:2002up}. This scheme uses a
$k_\perp$ measure to separate emissions into two phase space regions,
a low-$k_\perp$ region described by the parton showers and a
high-$k_\perp$ region described by the matrix elements. In order to
get a smooth transition between the two regions, the matrix element
emissions are treated as similarly as possible to parton shower
emissions. This is done by clustering the event using the $k_\perp$
jet clustering algorithm \cite{Catani:1991hj,Catani:1993hr} to find
its corresponding ``parton shower history'', \ie, the sequence of
parton emissions which would have been necessary for the parton shower
to generate the event. The event is then reweighted by using the
values of $k_\perp$ at the clustering nodes as scales in the running
$\alpha_s$ couplings (as is done in the parton shower), and a Sudakov
suppression factor is applied to get the probability for this
particular event to be generated without further emissions above the
$k_\perp$ cutoff scale. After that the event is showered, allowing
only shower emissions below the cutoff scale.

An alternative way to accomplish an equivalent result has been proposed by
Mangano~\cite{MLM}. In his approach, the multiparton events are
reweighted by $\alpha_s$ factors in the same way as in the CKKW
prescription, but no Sudakov reweighting is performed. Instead, the
event is showered, and then discarded if the showering generates
emissions harder than the phase space cutoff.

In the current MadGraph/MadEvent  release, both of these matching schemes are
implemented. For the MLM matching option, the showering and
vetoing of events with too large emissions are implemented in the
Pythia interface included in the Pythia-PGS package (see Section
\ref{sec:pythiapgs}), while the CKKW matching, including both $\alpha_s$ and
Sudakov reweighting, is so far implemented only at the matrix element
creation level.

In the original version of the MLM matching procedure, which is
implemented in Alpgen \cite{Mangano:2002ea}, the phase space separation is
defined in terms of cone jets. The MadGraph/MadEvent
implementation allows the user to choose whether to use cone jets or $k_T$
clustered jets, enabling a more immediate comparison to the
original CKKW formulation as implemented in Sherpa \cite{Gleisberg:2003xi}.

A comparison between event generators implementing matching
of matrix elements and parton showers is underway \cite{matchcomp}, including
MadGraph/MadEvent, Alpgen, Sherpa, HELAC~\cite{Kanaki:2000ey} and
Ariadne~\cite{Lonnblad:1992tz}. $W^\pm$+jets production 
at Tevatron and the LHC is there used as a case study.

It should also be noted that a similar matching method for matching of
MadEvent matrix elements to Pythia showers has been implemented
independently by Mrenna \cite{Mrenna:2003if} and has been successfully
used to describe Tevatron $W$+jets backgrounds.

\section{Tools}

The parton-level events generated with MadEvent are stored in the
so-called ``Les Houches accord event file
format''~\cite{Alwall:2006yp}.  Within 
this format enough information for each event is available so that
some of the data analysis usually performed during the event
generation phase, such as plotting, estimating PDF's errors or scale
variations can be deferred to a later stage. To this aim, simple
routines have been developed to perform some tasks ``off-line'', \ie,
directly on the event files produced.  The main reason is to improve
versatility and save time.  Generating events is a CPU expensive
activity, which, in some cases, can take many hours. Therefore, it is
not desirable to have to rerun codes only for making new plots or
switching from one scale choice to another.  Another important
advantage in working directly with the events is that the tools
developed are ``independent'' of how the events were generated and can
be used with any event set in the Les Houches format. In this respect,
the applications presented below can be used with events produced by
any matrix element generator.  The expert user is invited to develop
his/her own tools and make them available to the physics community.

\subsection{Analysis platforms: MadAnalysis and ExRootAnalysis}
\label{ssec:platforms}

Two platforms for performing studies on the event samples at all
stages of the simulations (parton, hadron and detector level) are
available. Here we describe them briefly.

MadAnalysis is a fortran-based analysis tool which allows you to
select events (by setting further acceptance or selection cuts) and
produce plots for events read in LHEF (parton level) and LHC Olympics
(detector level) formats.  Plots are simple ASCII files,
output either in a minimal format (just title and comments and $x y$ list)
which can be read by GnuPlot~\cite{gnuplot} with a simple script (provided), or in
TopDrawer~\cite{topdrawer} format.  A perl script to overlay plots of the corresponding
quantities for two event sets (or the same set at different stages of
the simulation chain) is also provided.

Particles, objects and reconstructed objects (like jets) in the final
state can be organized in classes, on which cuts and plots are based.
Particles or objects in each class are ordered with respect to a
one-particle quantity variable (the default is the transverse
momentum).  For instance, it is natural (at the parton level) to
define any light quark and gluon as a jet.  The user selects the cuts
and the details of the plots by editing a {\tt ma\_card.dat}.  Typical
quantities that can be plotted are one-particle quantities, such as
transverse momentum ($p^T$) and the rapidity ($y$) or pseudo-rapidity
($\eta$), or multi-particle quantities, such as two-body or three-body
invariant masses ($m(i,j),m(i,j,k)$), or "distances" (such as $R(i,j)$
or $k_T(i,j)$)) for each pair of final-state particles. A rather
exhaustive library of kinematic functions is available to the user, as
well as the possibility of defining new functions of the momenta of
any number of particles in a very user-friendly way.

The {\Analysis} package allows to store and analyze events in a
{\tree} format~\cite{root}. Normally, {\ROOT} allows to store data in
several different formats. However, there are some crucial
differences, which makes the {\tree} format more attractive than
others:
\begin{itemize}
  \item information is stored in arrays of objects, which enables
efficient storage (compression using the {\ROOT} gzip algorithm) and
retrieval; 
  \item the same C++ classes are used for creating the {\tree} and for
analysing the stored data. 
\end{itemize}

{\tree} objects are created from particles generated by MadGraph and
Pythia or objects produced by PGS (in most of the cases physics
objects like jets, electrons, {\etc}), in order to perform analysis in
a {\ROOT} environment.

The {\Analysis} package can be subdivided into several subsystems:
basic framework of few classes providing event loop, event selection
and basic operations with a {\tree} file; modules selecting events and
objects to be analysed at per event and per object level; and modules
analyzing selected events.

For example, a selector module can select and group partons generated
by MadGraph into several classes (such as leptons, jets, top quarks,
{\etc}) according to their status and particle identification
number. Any number of classes can be specified in the configuration
file. After the partons have been classified, an analysis module can be
used to produce series of standard plots for each class of partons.

Documentation on the content of the {\tree} is available on the
web~\cite{tree_structure}.

\subsection{Decaying unstable particles in the SM and beyond}

MadEvent includes a tool, Decay, that performs the decays
of unstable particles in the Standard Model directly on the
parton-level events (weighted or unweighted). At present, a total of
68 decay modes are included for $\tau,W,Z,t$  and $h$ decays.  The advantage
of using Decay is obvious: from the point of view of MG/ME the
generation of events is faster for a simple final state. When a
detailed knowledge of the spin-correlations is not needed, 
Decay, which is only keeping the diagonal terms in the
spin-correlation matrix in the helicity basis, is an accurate and very
efficient tool to get a multi-particle final state.  
A generalization and considerable improvement of Decay 
has been recently released by Meade and Reece~\cite{Meade:2007js},
under the name of BRIDGE. BRIDGE can take an arbitrary model as input 
and compute the widths (and hence branching ratios) for all the kinematically 
allowed two-body and  three-body tree level decays. 
It can also decay generated events, choosing decay 
modes randomly according to the branching ratios.

\subsection{Hadronization and detector simulation using Pythia and PGS}
\label{sec:pythiapgs}

One objective in the latest development of MadGraph/MadEvent has been
to facilitate hadronization and detector simulation. We have therefore
made a package available which includes the hadronization package
Pythia 6.4 \cite{Sjostrand:2006za} and the fast detector simulation
package PGS 4 \cite{PGS4}, as well as the parton density package
LHAPDF \cite{Whalley:2005nh}, a stripped version of CERNLIB, and the
utility STDHEP \cite{STDHEP} which is used for communication between
Pythia and PGS. The package also includes main programs for running
Pythia and PGS, and an interface for reading MadEvent files into
Pythia. This package is available for download, or can be used
directly in the on-line event generation to simulate fully hadronized
and detector reconstructed events.

The MadEvent-Pythia interface reads the Les Houches Event file output
of MadEvent, and communicates model parameters such as particle masses
to Pythia. The behaviour of Pythia is determined by the input file
{\tt pythia\_card.dat}. In the case where it finds particles in the
events which are considered massless in MadEvent but massive in
Pythia, such as electrons or muons, the particles are given their
Pythia mass, and momenta are redistributed to account for the
changes. The interface also allows for matching of jet production by
Pythia parton showers with multiparton samples generated by MadEvent,
see Section \ref{sec:matching}.

The outputs of the Pythia main program are a binary file with the full
hadronic event information in the STDHEP output format and a text file
in the Les Houches Event file format with information on resonances,
jets, leptons, and missing energy. Jet clustering for the latter file
is performed either using a $k_T$-clustering routine or a cone jet
routine.

The PGS main program reads the STDHEP file from the Pythia generation,
and performs detector simulation using the detector setup given in the
input file {\tt pgs\_card.dat}. The output is a text file in the PGS4
LHC Olympics format with information on the trigger status of the
event, jets, leptons and missing energy.

All the output files can be read by the ExRootAnalysis package to get
ROOT files and plots for the different stages. They can also be read
by the MadAnalysis package for simple analysis purposes (see Section
\ref{ssec:platforms}). The events are consistently numbered in the
output files throughout the chain.

\section{Applications to hadron collider physics}
\label{sec:applications}

In this section we present several examples of the types of studies
that can be performed with the MadGraph/MadEvent package. The aim is
to show how easily the various new features can be used to perform
signal and background analyses with both theoretical and experimental
aims. Even though the discussion is kept concise for space reasons,
we
stress that all the results obtained in the following could stand in a
dedicated publication and that some of them are original and presented
here for the first time.

\subsection{Higgs search in $pp \to h \to W^+W^-$: signal and backgrounds}

For a Higgs boson of moderate mass, $140$ GeV $< m_h < 170$ GeV, the golden
discovery channel is via gluon-gluon production and successive decay
into a pair of leptonically decaying $W$ bosons,
Fig.~\ref{fig:hww_diag}.
\FIGURE[t]{
\epsfig{file=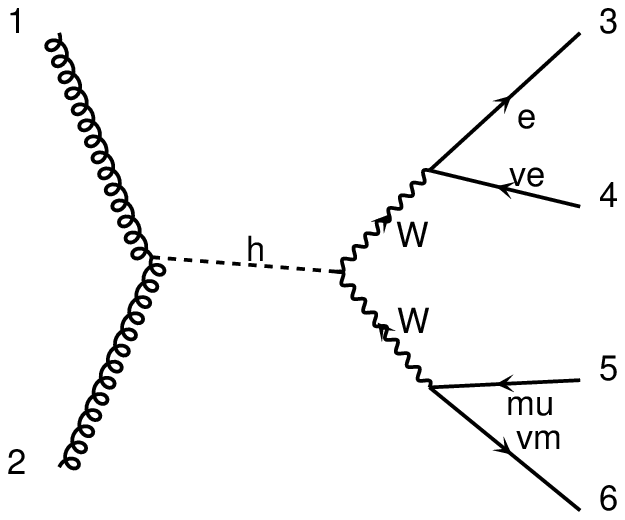, width=5cm}
\caption{The diagram corresponding to the process 
$pp \to h \to W^+W^-\to e^- \mu^+ \bar \nu_e  \nu_\mu$}
\label{fig:hww_diag}
} This process has a large cross section and
also special kinematics characteristics, notably the angular distance between
the leptons coming from the $W$'s decays, which can be exploited to
improve the signal over background ratio. Reducible backgrounds are
dominated by processes which involve single top and $t\bar t$
production, which typically yield a larger number of jets in the final
state and can be therefore suppressed by imposing a jet veto. Many
studies have been performed on this channel and its backgrounds both
at the experimental and the Monte Carlo level, sometimes with
different generators, selection cuts and
approximations~\cite{Buttar:2006zd}.
 
In the following we discuss how such a study can be
done using MG/ME. Our aim is to show how easy it is to generate all
the necessary event samples in one go, from parton-level to detector
objects, while keeping all the relevant non-trivial features of the
signal (such as spin correlations) and the background (resonant and
non-resonant contributions). In particular, the example makes use of 
the following features:

\begin{itemize}
\item Inclusion of more than one process in the MadGraph
generation (see Fig. \ref{fig:ww-card}). The different processes
included in this way are then produced in proportion to their cross
section, so the event file produced automatically gets the correct mix
of unweighted events.
\item The possibility to generate the backgrounds corresponding to 
a given final signature (in this case $e^\pm \mu^\mp b \bar b$), by
including all the classes of diagrams (including resonant and
non-resonant diagrams) consistently at once. The information on the
relative contributions from the various resonances is available on
event-by-event basis.
\item The interfaces to Pythia for decay and hadronization, PGS
for detector simulation and MadAnalysis for plotting, analyzing and comparing
observables  at the various stages of the simulation.
\end{itemize}

\begin{figure}
\footnotesize{
\begin{verbatim}
# Begin PROCESS # This is TAG. Do not modify this line

pp>h> e- mu+ ve~ vm  @1 # Signal Process 
QCD=0                   # max qcd order
QED=4                   # max qed order
HIG=1                   # max hig coupling order
end_coup                # the coupling list is over

pp > e- mu+ ve~ vm/h @2 # Irriducible background pp > W+W-> e- mu+ ve~ vm
QCD=0                   # max qcd order
QED=4                   # max qed order
end_coup                # the coupling list is over

pp > bb~e-mu+ve~vm/h @3 # Reducible background non-, single-, double-resonant
QCD=2                   # max qcd order
QED=4                   # max qed order
end_coup                # the coupling list is over

done                    # the process list is over

# End PROCESS  # This is TAG. Do not modify this line
# Begin MODEL  # This is TAG. Do not modify this line
heft
# End   MODEL  # This is TAG. Do not modify this line
# Begin MULTIPARTICLES # This is TAG. Do not modify this line
p  uu~dd~ss~cc~g 
# End  MULTIPARTICLES # This is TAG. Do not modify this line
\end{verbatim}}
\caption{\label{fig:ww-card} The {\tt proc\_card.dat} used for signal
$pp \to h \to W^+W^-\to e^- \mu^+ \bar \nu_e \nu_\mu$ and the
irreducible and reducible backgrounds. By a careful choice of the
couplings and by vetoing the Higgs as an intermediate particle, the
backgrounds are correctly included. The third process include
non-resonant, single-top resonant and $t\bar t$ double-resonant
production.}
\end{figure}

We consider that both $W$'s decay leptonically, and for simplicity we
restrict to the different flavor case, $e^\pm \mu^\mp$.  As our
purpose is illustrative we use (a simplified version of) the selection
cuts used in Ref.~\cite{Davatz:2004zg}.  we choose the angular
distance between the two leptons in the transverse plane,
$\Delta\phi$, as the discriminating variable. As it is well-known,
this variable encodes the fact that in the signal the two leptons tend
to be produced close in phase space, due to the constraints coming
from angular momentum conservation and the purely left-handed
couplings of the $W$ to the leptons.

\
\FIGURE[t]{
\epsfig{file=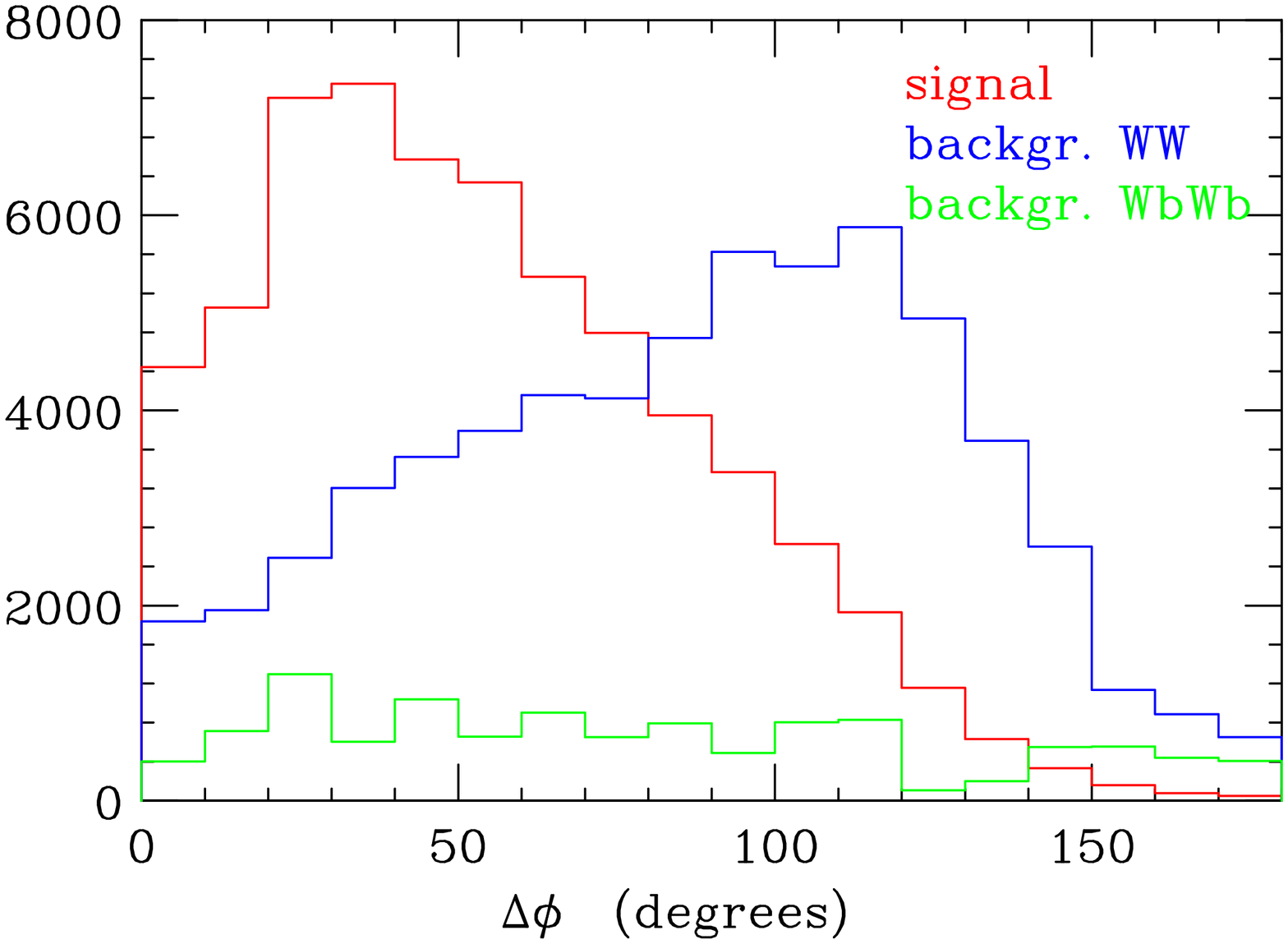, angle=90 ,width=7cm}
\hspace*{.5cm}
\epsfig{file=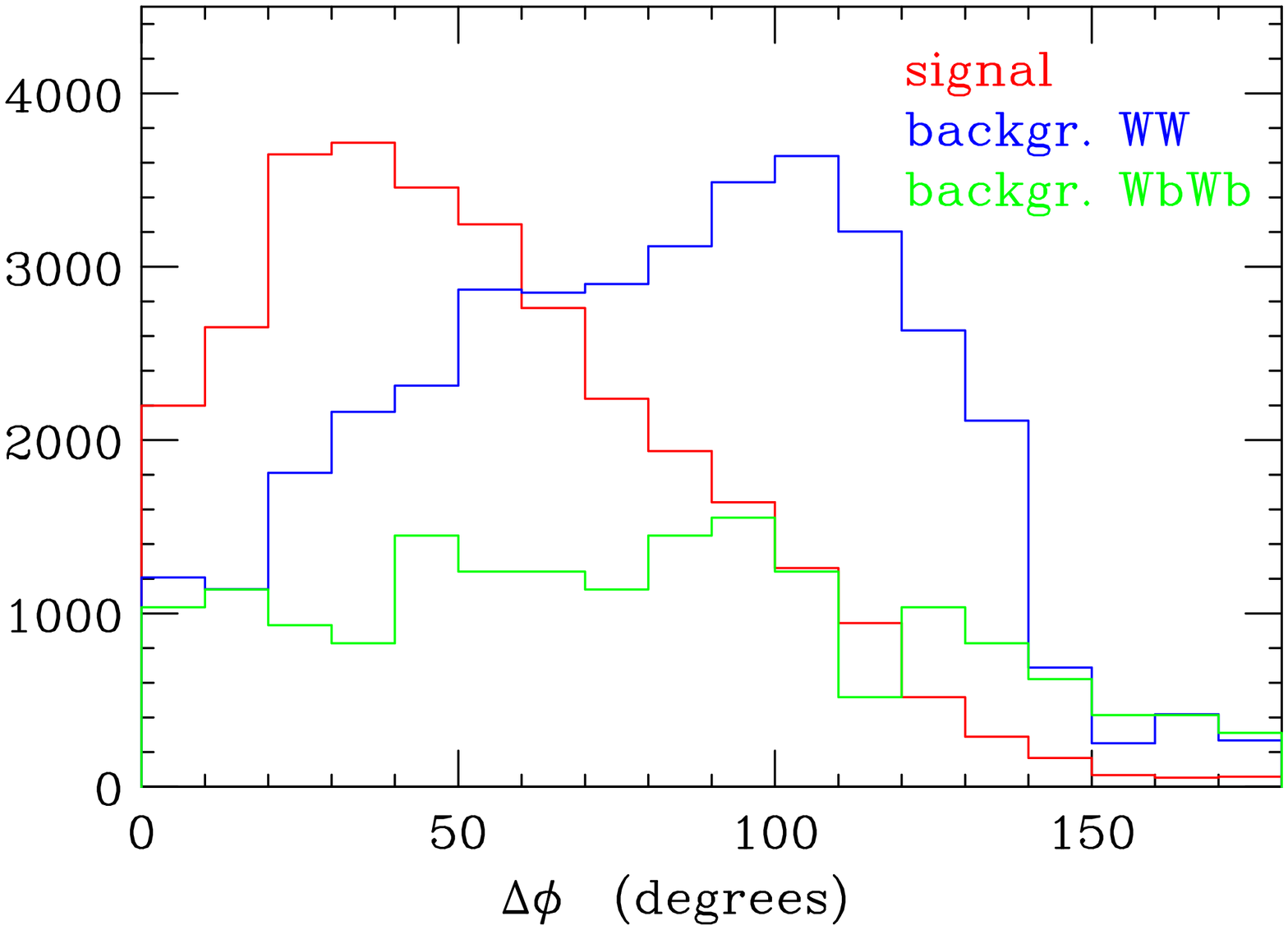   , angle=90, width=7cm}
\caption{$\Delta{\phi}$ distributions of the two leptons in $W^+W^-$
events at the parton (left) and hadron (right) level.}
\label{fig:hww_plots}
}

The selection cuts used in the $\Delta\phi$ distributions at the
parton and PGS level shown in Fig.~\ref{fig:hww_plots} are:
\begin{itemize}
\item leptons: $25$ GeV $<p_T^\ell< 50$ GeV, $|\eta^\ell|<2.4$,
      $\Delta R(\ell,\ell')>0.4$, and
$m(\ell,\ell')<80$ GeV;
\item missing $E_T$: $E_T^{\rm miss}>30$ GeV;
\item jet veto: no jets with $p_T^j>30$ GeV and $|\eta^j|<3$.
\end{itemize}

We note that there is not much difference in the $\Delta\phi$
distributions in going from the parton-level events (where the Higgs
has no transverse momentum and the two $W$'s might have a very small
one) to the fully simulated case. However, by plotting the $p_T$ of
the leptons we checked that the initial state radiation has a
non-negligible impact on the distributions, moving the average $p_T$
of the leptons considerably.

Another very interesting feature to study is how the relative
contributions coming from $WbWb$ final states with top-quark
non-resonant, single-resonant and double-resonant diagrams change as
the jet veto is applied.  This is a very good examples of a MC
generation that has to be done carefully~\cite{Kauer:2004fg}. In fact
single-resonant contributions are completely negligible and $WbWb$
production is completely dominated by $t\bar t$ in generic areas of
the phase space. On the other hand, the request of a jet veto
dramatically enhances the non-resonant and single-resonant
contributions which account for more than 50\% of the final $WbWb$
background.

\subsection{Higgs CP properties: $pp \to h jj$ in the HEFT}

In this example we show how the Higgs effective theory implementation
can be used to study the CP properties of a Higgs boson by looking at
the angular distributions of the jets in the process $pp\to
hjj$~\cite{Hankele:2006ma,Klamke:2007cu}.  Our study is done at the
parton level, but could be promoted to the hadron level without
effort.

We assume that the Higgs boson couples mainly to heavy quarks. In this case 
the Higgs boson will then be mainly produced by gluon fusion through 
a top quark loop. As discussed in Sec.~\ref{heft}, 
for Higgs boson masses smaller than two times the top quark mass, 
$m_h\lesssim 2m_t$, we can send the mass of the top quark in the loop 
to infinity $m_t\to\infty$, to a very good approximation. 
Effectively, this means that we contract the top quark loops in the Feynman
diagrams and get effective $ggh$, $gggh$ and $ggggh$ vertices.
In Fig. \ref{fig:hjj} three
diagrams with these effective vertices
contributing to $hjj$ production are presented.
These vertices are implemented into the HEFT model of MadGraph,
both for scalar and pseudo-scalar Higgs bosons.

\begin{figure}[t]
  \begin{center}
    \epsfig{file=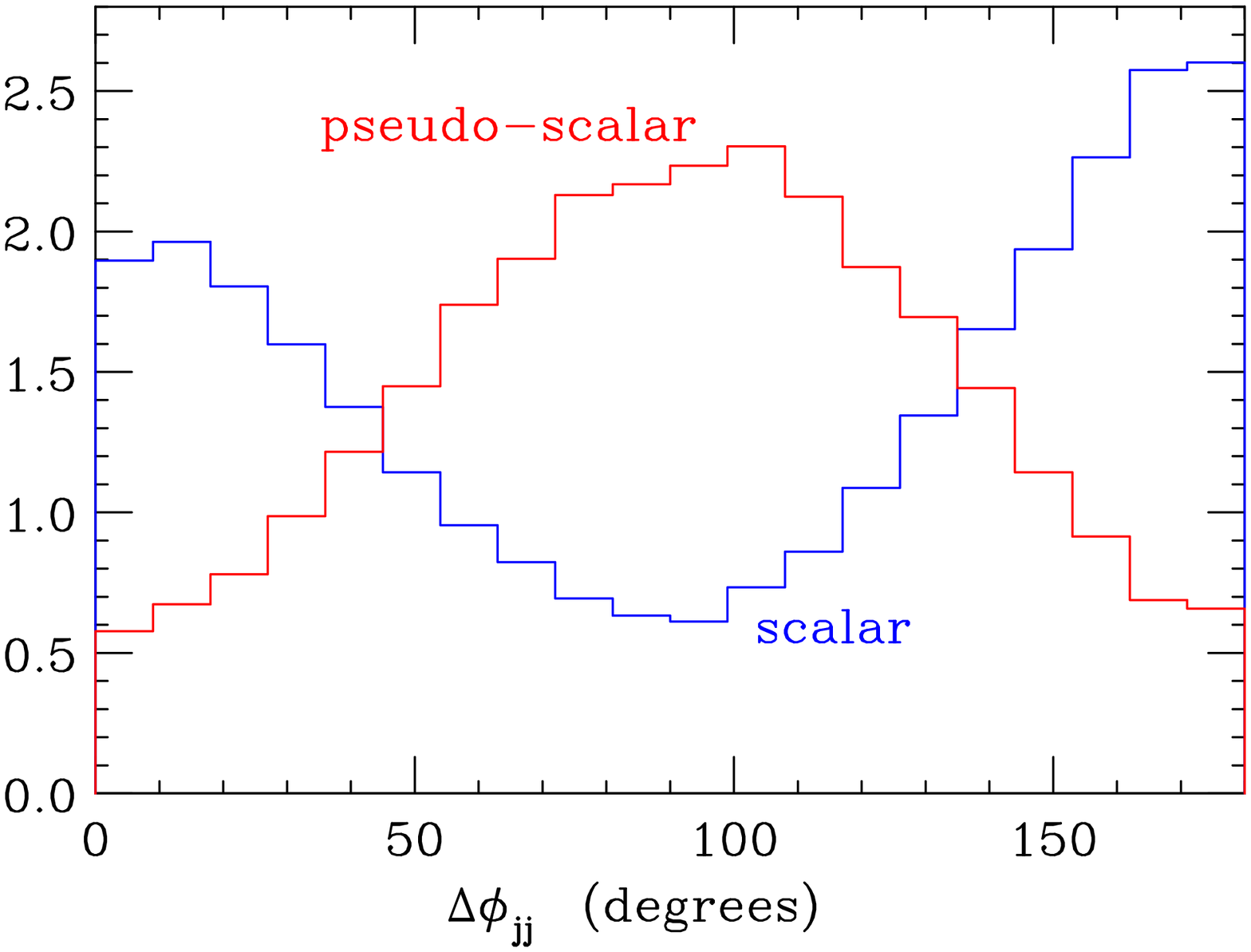,angle=90, width=0.6\textwidth}
  \end{center}
  \vspace{-15pt}
  \caption[]{$\Delta\phi_{jj}$ distribution for $pp\to hjj$ production
for a scalar and a pseudo-scalar Higgs boson through gluon fusion.
The plots are normalized. Cuts are as in \eqref{cuts}, $M_h= 120$ GeV.}
  \label{CP-plot}
\end{figure}

To investigate the CP properties of the Higgs boson we look at the
angle $\Delta\phi_{jj}$ \cite{Plehn:2001nj}, \ie, the angle between the transverse momenta
of the two jets.  We ask for one very forward and one very backward
jet by applying the following cuts on the jets:
\begin{equation}\label{cuts}
p_T(j) > 20 \textrm{ GeV},\qquad \Delta R_{jj}>0.4,\qquad
|\eta_{j_1}-\eta_{j_2}|>4,\qquad  \eta_{j_1}\cdot \eta_{j_2}<0.
\end{equation}
These cuts lead to a signal comparable with Higgs boson production
through W boson fusion. In Fig.~\ref{CP-plot}, $\Delta\phi_{jj}$ is
plotted with these cuts for a Higgs boson mass of $M_h=120$ GeV, for
the pure scalar and pure pseudo-scalar cases.


\FIGURE[t]{
\epsfig{file=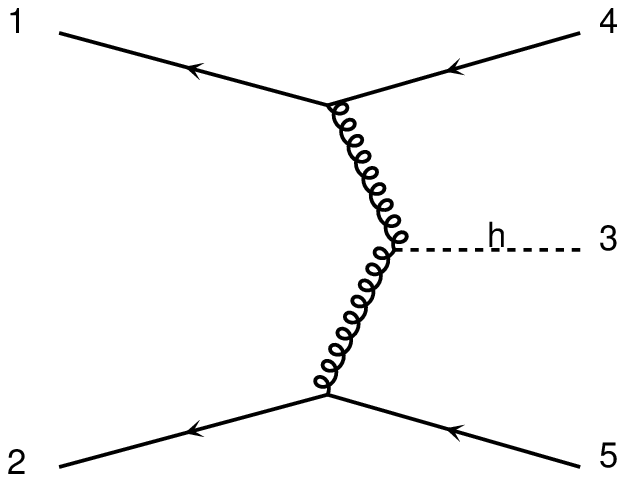, width=4cm}\hspace*{.3cm}
\epsfig{file=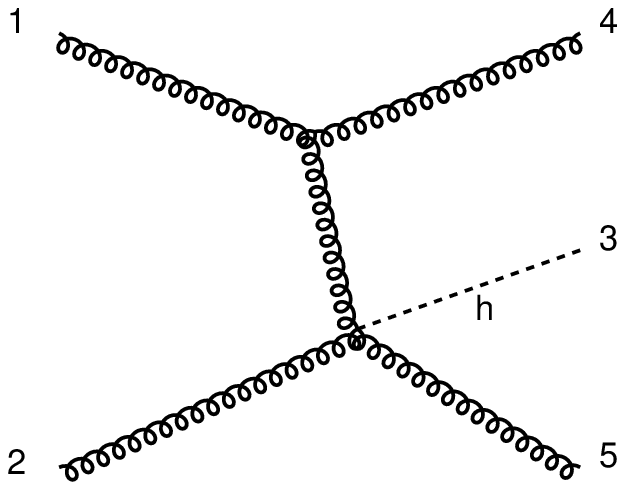, width=4cm}\hspace*{.3cm}
\epsfig{file=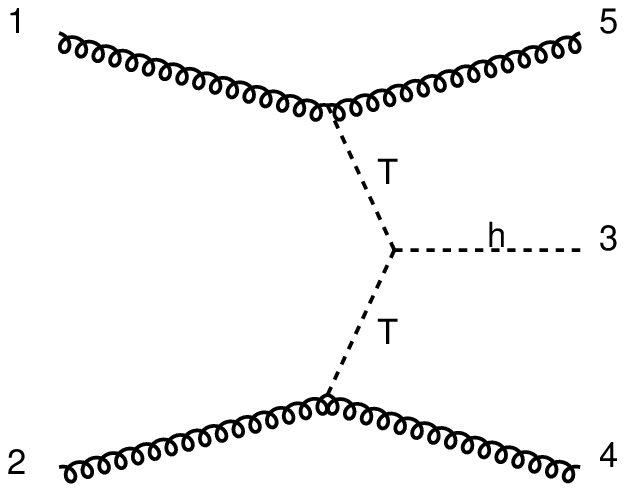, width=4cm}
\caption{Selection of the diagrams contributing to $hjj$ production.
The dashed line correpsonds to the auxiliary tensor particle $T$.
For a pseudo-scalar Higgs the diagram on the \emph{rhs} does not contribute.}
\label{fig:hjj}
}

\subsection{Spin of a new resonance from lepton angular distributions}

Here we present the angular distributions for different intermediate
particles in the process $p\bar{p}\to X\to \mu^+\mu^-$,
where $X$ is a $s$-channel resonance. We consider the three cases
where $X$ is a spin-0, spin-1 or a spin-2 particle.

This study exploits the possibilities of the user model framework to
introduce new particles and interactions and also makes use of the
full simulation chain, from parton level events to detector
reconstruction.

To minimize the effects of a non-zero transverse momentum of the
intermediate state $X$, we use the angle $\theta$ introduced by
Collins and Soper to study spin correlations~\cite{Collins:1977iv}.
The angle $\theta$ is defined as follows. Let $p_A$ and $p_B$ the
momenta of the incoming hadrons in the rest frame of the muon pair. If
the transverse momentum of the muon pair is non-zero, then $p_A$ and
$p_B$ are not collinear. Let us define an axis in such a way that it
bisects the angle between $p_A$ and $p_B$. The angle $\theta$ is
defined to be the angle between this axis and the $\mu^+$ momentum in
the muon pair rest frame.

For (leading-order) parton level results, where we do not include
extra radiation, the transverse momentum of the muon pair is zero.
Then the angle $\theta$ is the same as the more commonly used angle
$\theta^{\star}$, \ie, the angle between the $\mu^+$ momentum in the
muon pair rest frame and the beam direction in the lab frame.

\begin{figure}[h]
  \hfill
  \begin{minipage}[t]{.32\textwidth}
    \begin{center}
      \epsfig{file=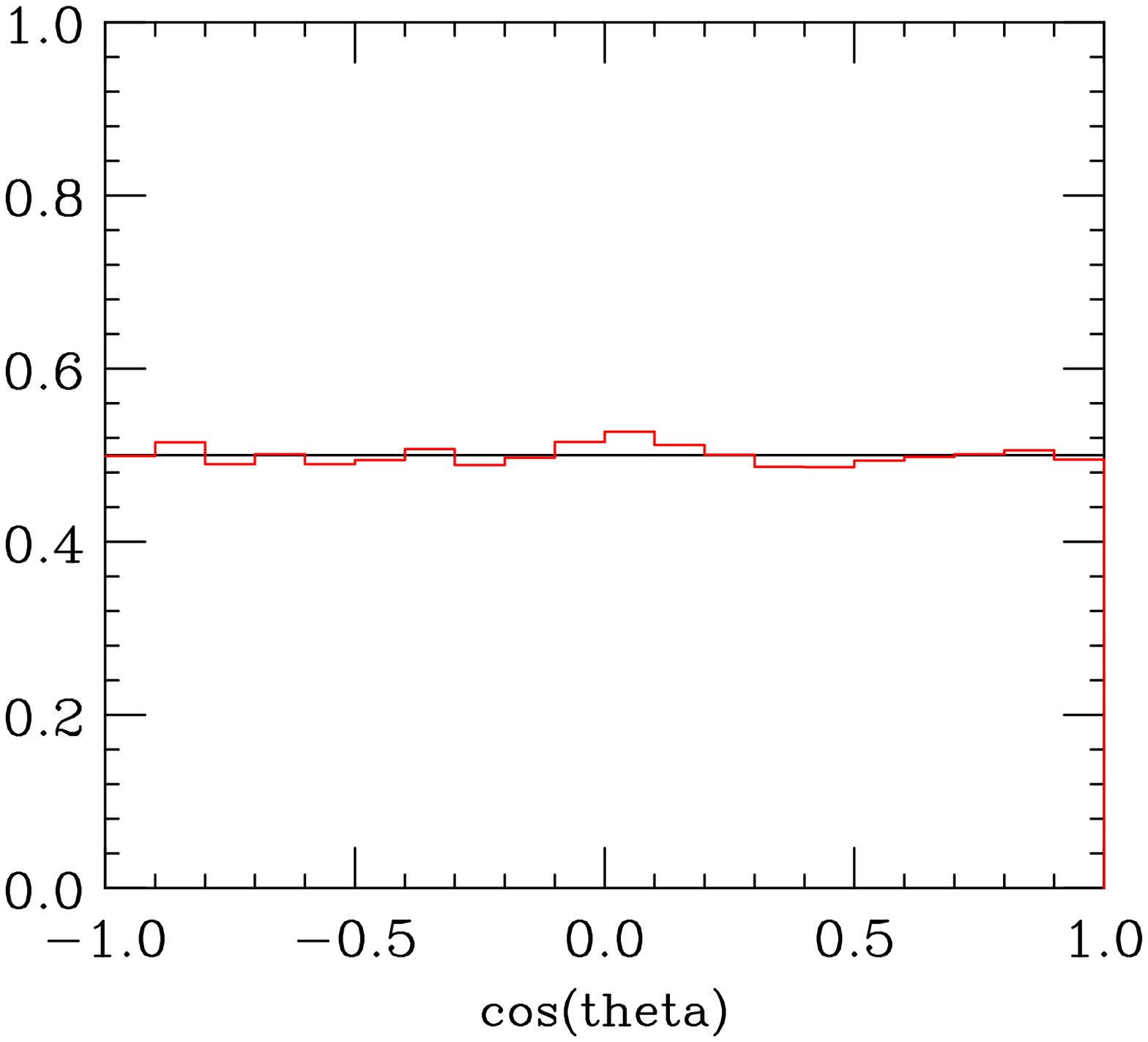,angle=90, width=1\textwidth}
    \end{center}
  \end{minipage}
  \hfill
  \begin{minipage}[t]{.32\textwidth}
    \begin{center}
      \epsfig{file=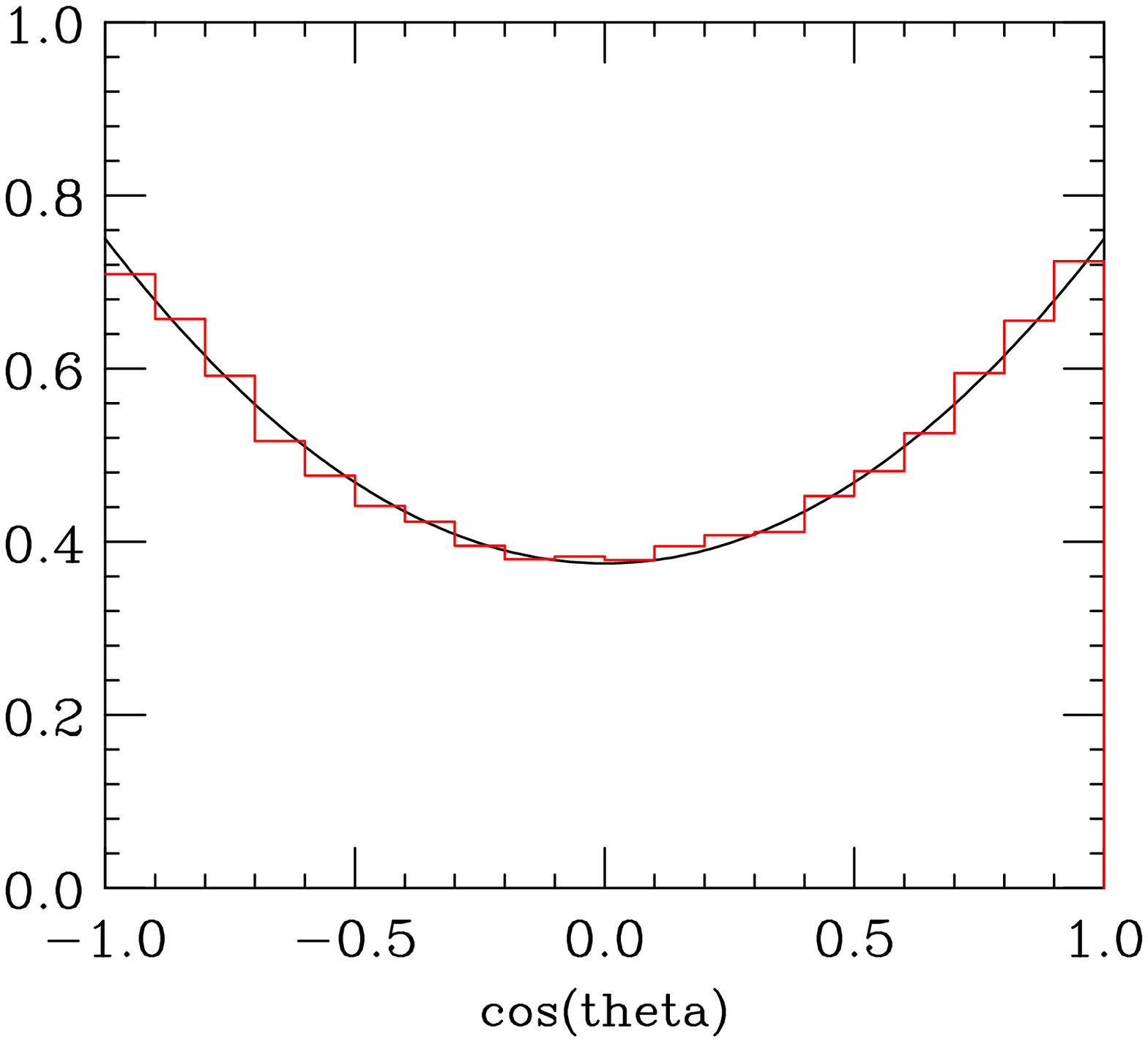, angle=90,width=1\textwidth}
   \end{center}
  \end{minipage}
  \hfill
  \begin{minipage}[t]{.32\textwidth}
    \begin{center}
      \epsfig{file=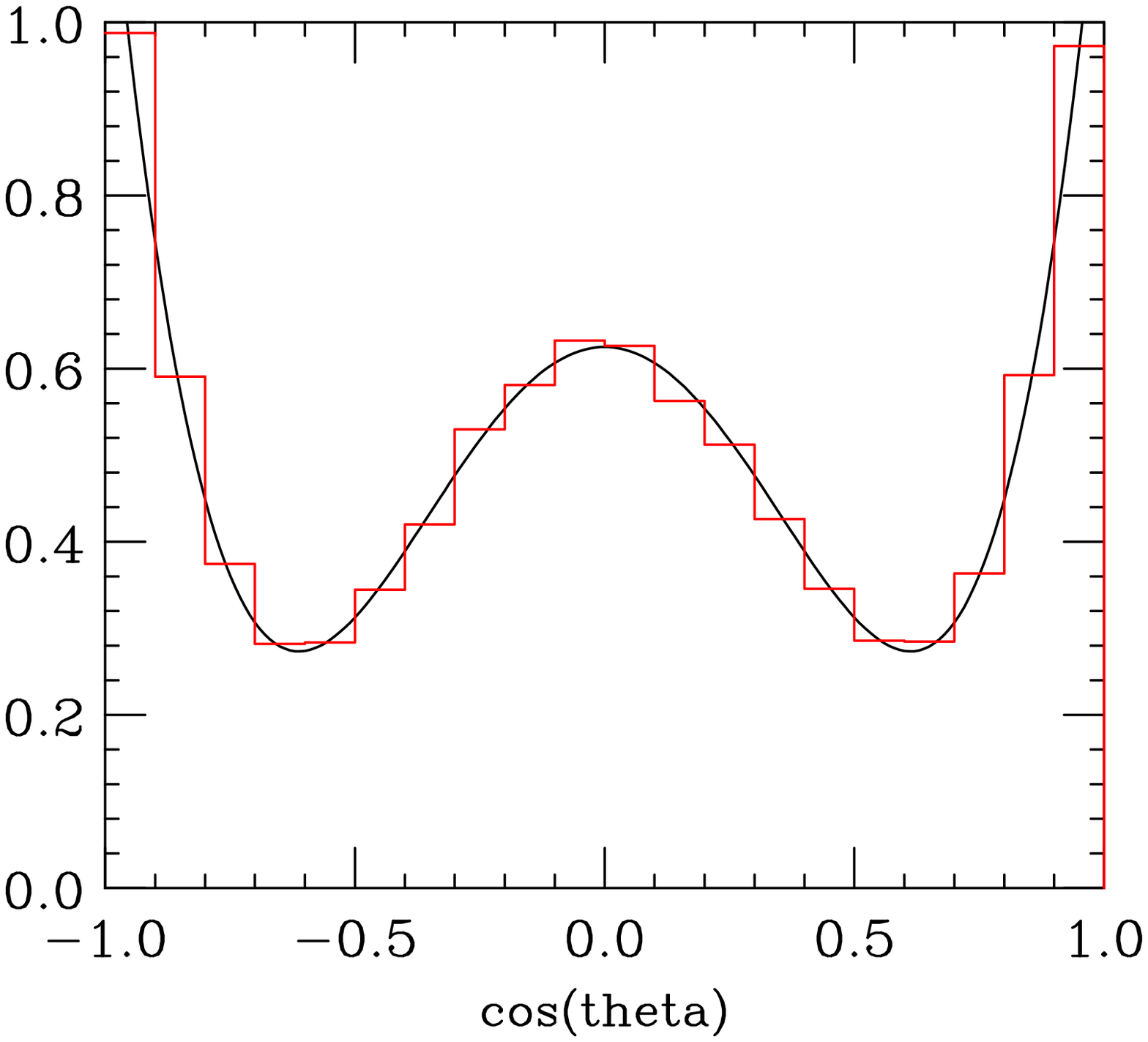, angle=90,width=1\textwidth}
   \end{center}
  \end{minipage}
  \hfill
      \caption{The normalized cross section as a function of $\cos\theta$
in $q\bar{q}\to X\to \mu^+\mu^-$. 
\emph{Left} for a spin-0 particle, 
\emph{center} for a spin-1 particle,
\emph{right} for a spin-2 particle. No cuts applied.}
      \label{distributions}
\end{figure}

In Fig.~\ref{distributions} the cosine of this angle is plotted for the
three different cases of the spin of the particle $X$. This figure includes
only $q\overline{q}$ initial states. 

\begin{figure}[h]
  \hfill
  \begin{minipage}[t]{.48\textwidth}
    \begin{center}
      \epsfig{file=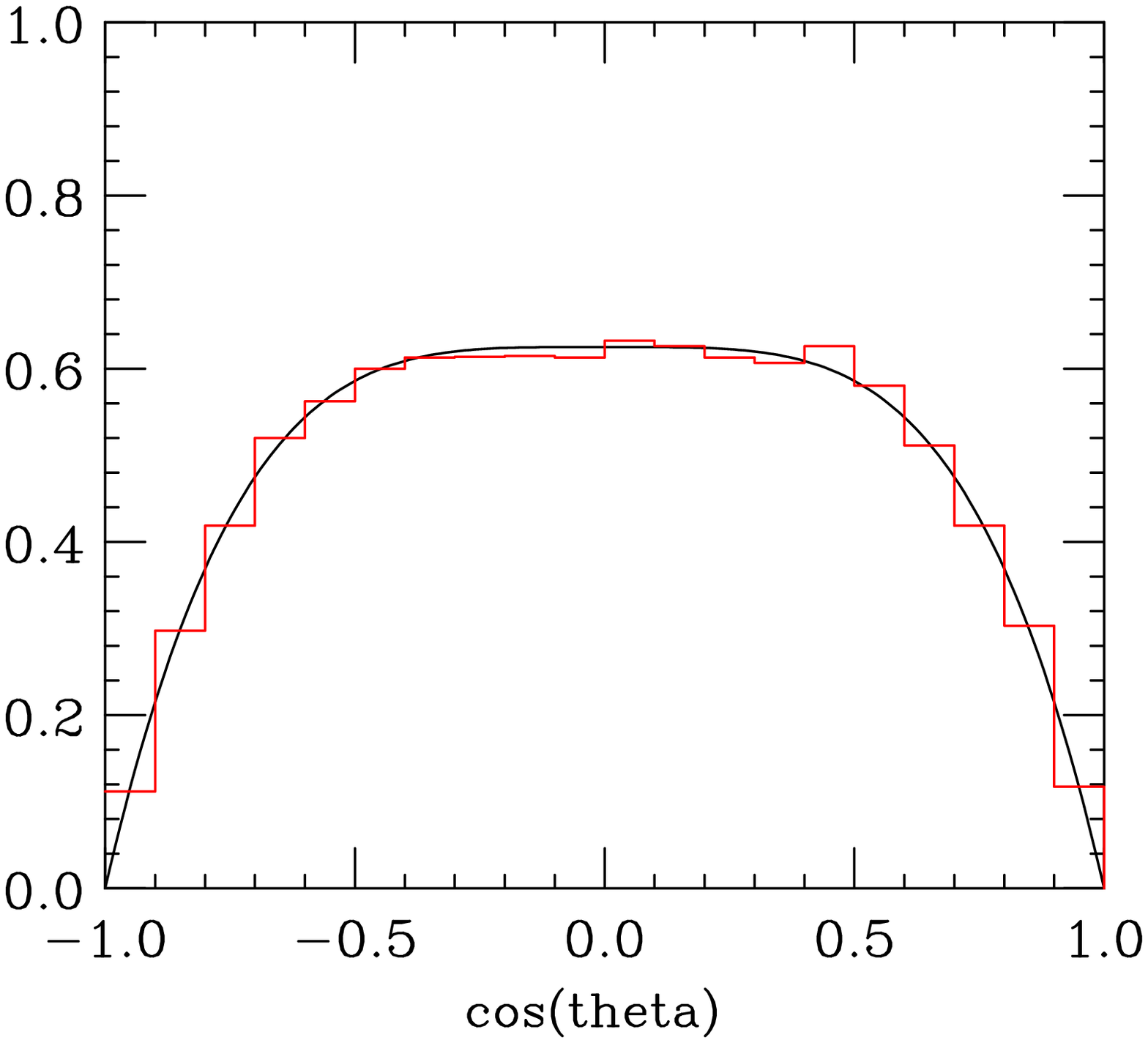,angle=90, width=0.8\textwidth}
    \end{center}
  \end{minipage}
  \hfill
  \begin{minipage}[t]{.48\textwidth}
    \begin{center}
      \epsfig{file=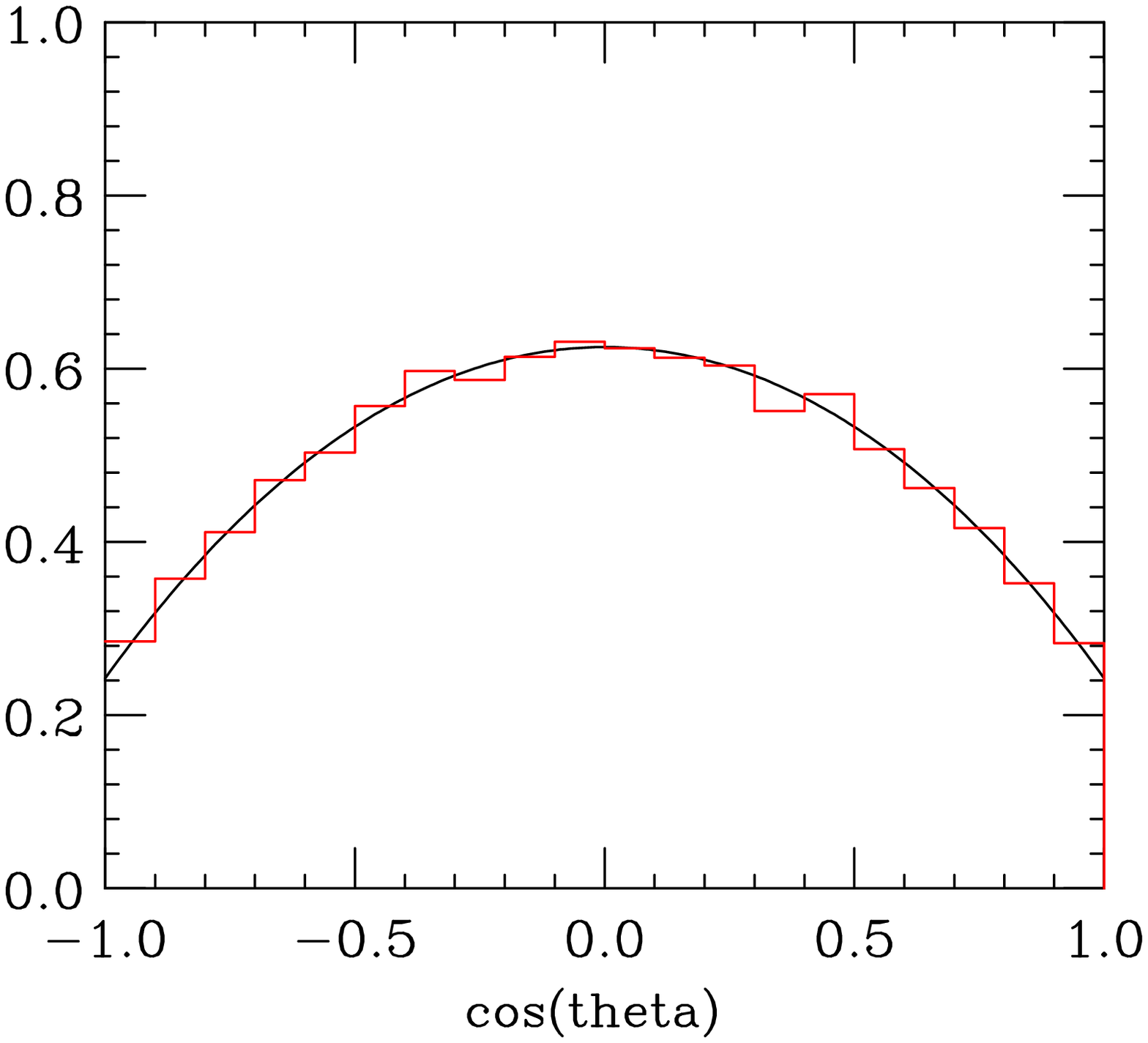, angle=90,width=0.8\textwidth}
   \end{center}
  \end{minipage}
  \hfill
\caption{The normalized cross section as a function of $\cos\theta$
in $gg\to\,\textrm{spin-2}\,\to \mu^+\mu^-$ (\emph{left})
and $pp\to\,\textrm{spin-2}\,\to \mu^+\mu^-$ (\emph{right}).
The mass of spin-2 particle is 1 TeV. No cuts applied.}
\label{spin_gg}
\end{figure}

The spin-2 particle can also be created by gluon fusion, which
dramatically impacts the distribution. The angular
distributions of the muons depend, in general, on the way the $X$ 
resonance was produced. In Fig.~\ref{spin_gg} the muons angular distribution
is plotted for the gluon initial state and for the 
sum of the gluon and quark initial states, for a spin-2 resonance
with a mass of 1 TeV.


\begin{figure}[h]
  \hfill
  \begin{minipage}[t]{.32\textwidth}
    \begin{center}
      \epsfig{file=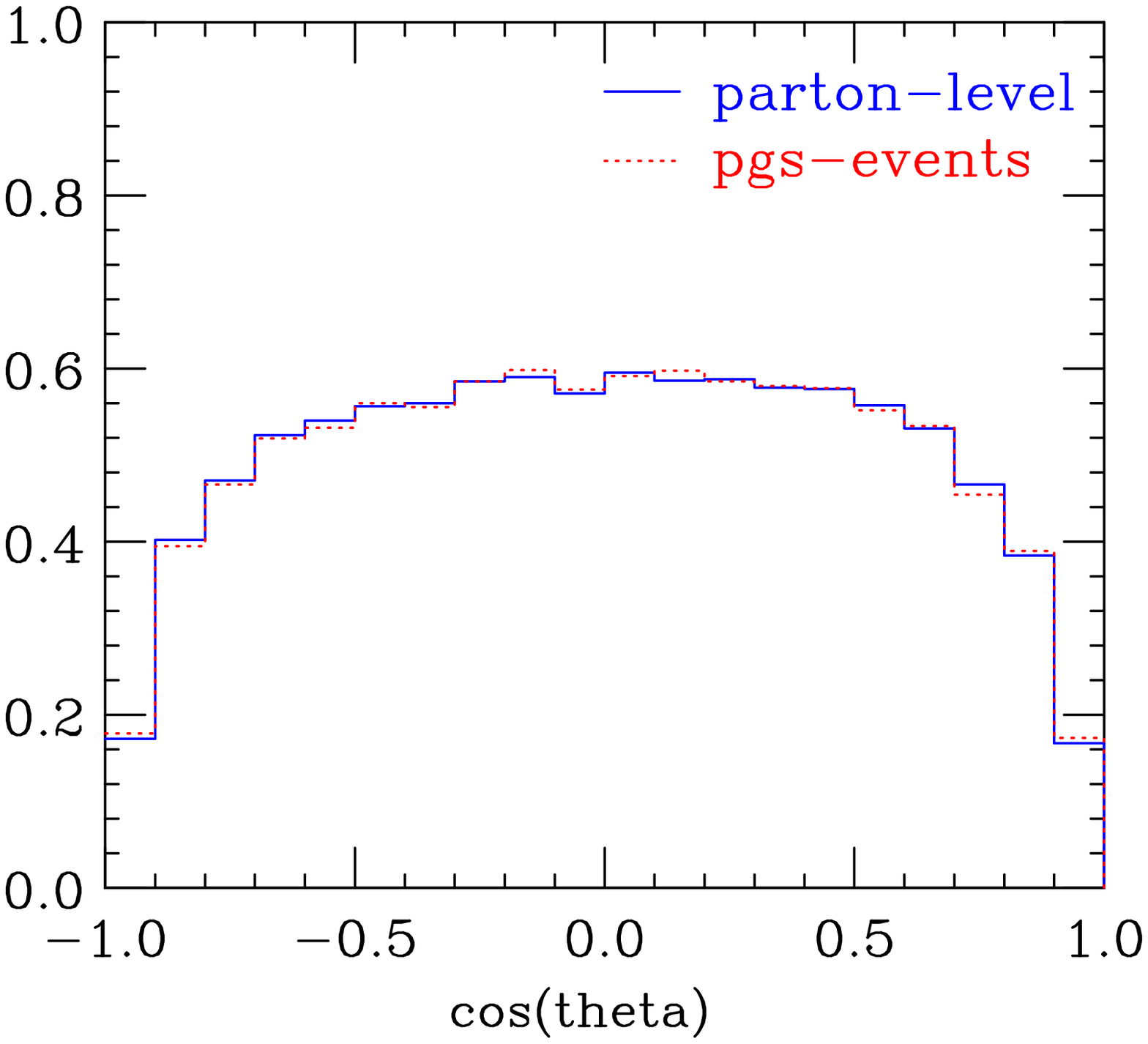,angle=90, width=1\textwidth}
    \end{center}
  \end{minipage}
  \hfill
  \begin{minipage}[t]{.32\textwidth}
    \begin{center}
      \epsfig{file=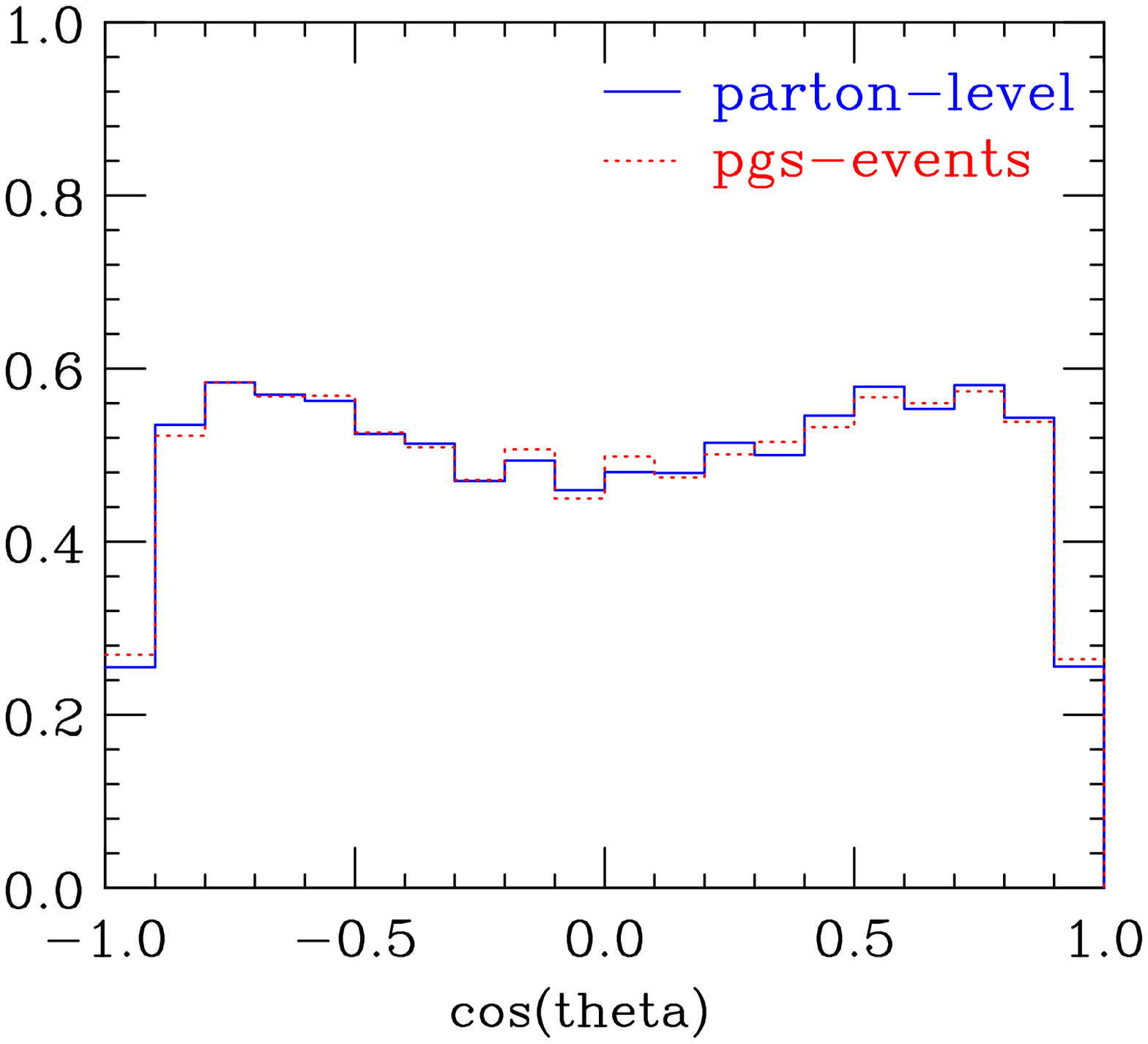, angle=90,width=1\textwidth}
   \end{center}
  \end{minipage}
  \hfill
  \begin{minipage}[t]{.32\textwidth}
    \begin{center}
      \epsfig{file=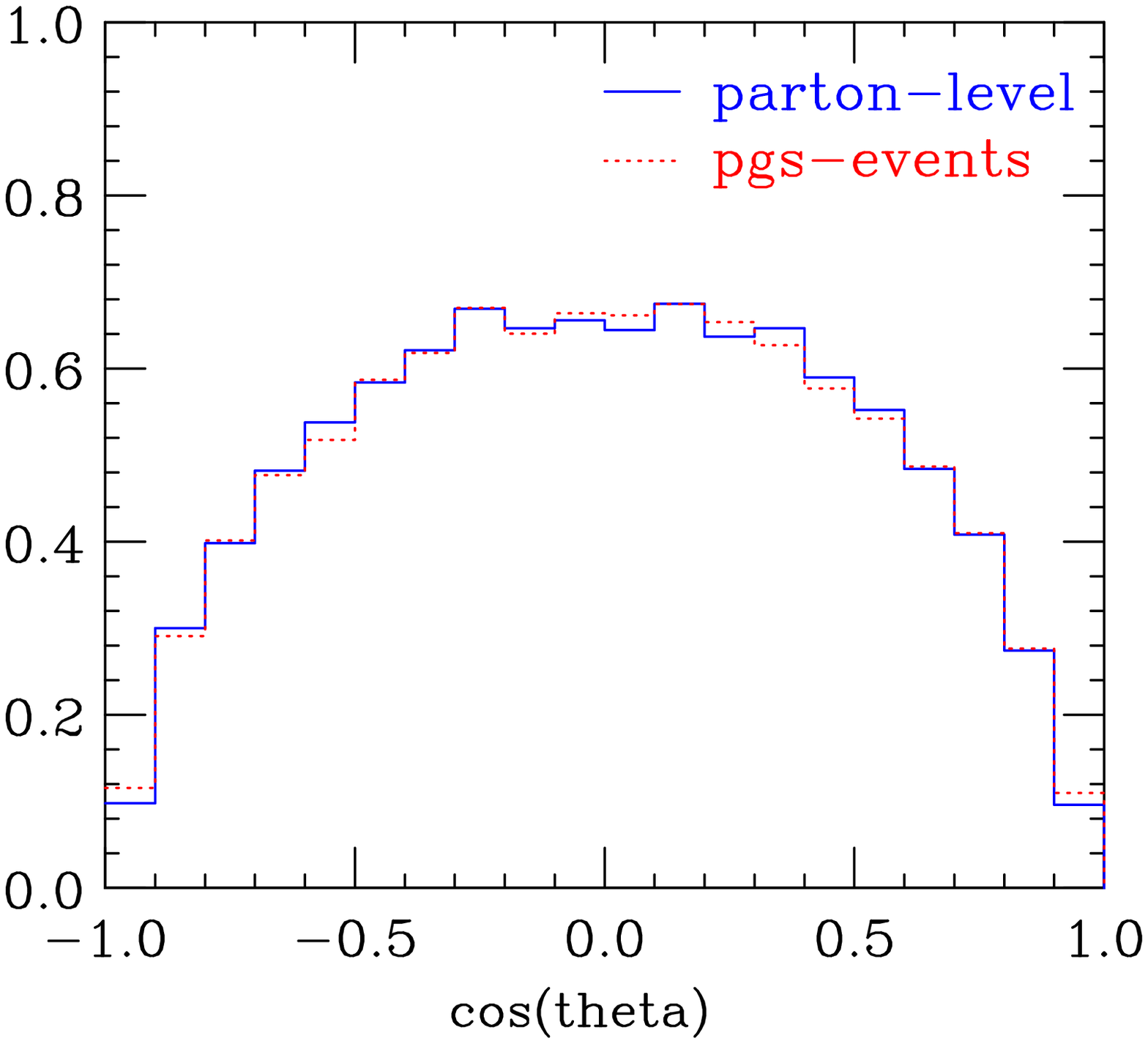, angle=90,width=1\textwidth}
   \end{center}
  \end{minipage}
  \hfill
      \caption{The normalized cross section as a function of $\cos\theta$
in $pp\to X\to \mu^+\mu^-$ for the reconstructed events after
simulation of extra radiation using Pythia and detector response
using PGS (\emph{dotted, red}) and 
parton level events with rapidity cut $|\eta_{\mu^+,\mu^-}|<2.4$
(\emph{solid, blue}).
\emph{Left} for a spin-0 particle, 
\emph{center} for a spin-1 particle,
\emph{right} for a spin-2 particle. Mass of the $X$ particle is 1 TeV.}
      \label{spin-pgs}
\end{figure}

Including initial state radiation, showering and hadronization,
 does not modify the lepton distributions significantly. If we use 
PGS to simulate detector response we can get relatively reliable results
for the CMS experiment at the LHC. In Fig.~\ref{spin-pgs}, reconstructed
events generated with PGS are plotted.
In the same figure  the parton level
results are plotted with an acceptance cut for the rapidity of the muons
$|\eta_{\mu^+,\mu^-}|<2.4$. Due to this cut the events
with $\theta\approx 0$ (\ie, both muons in the same forward direction)
and $\theta\approx \pi$ (\ie, muons in opposite forward-backward directions)
are not detected. We see that for a 1 TeV resonance the lepton
distributions hardly change at all.

\subsection{Single-top associated Higgs production in a generic 2HDM}

In this example we make use of the 2HDM implementation in
MadGraph/MadEvent and of the associated calculator.

The $t\overline{t}$ associated Higgs boson production is well known to
have a non-negligible cross section at LHC ($\simeq 1$ pb) for
moderate values of the Higgs mass ($m_h\simeq 100$ GeV). This can be
estimated starting from the large $t\overline{t}$ pair production
cross section and then accounting for the large top-Higgs Yukawa
coupling and phase space suppression. Since the single top production,
in particular in the $t$-channel, also has a sizeable rate at LHC
(more or less a third of the $t\overline{t}$ production) it could
naively be expected that the single-top associated production of the
Higgs boson would be promising. It has been shown, however, that the
cross section for single top in association with the Higgs is sizeably
smaller than expected~\cite{Stirling:1992fx,Maltoni:2001hu}.

The single-top associated production of the SM Higgs boson appears to
be of the order of $100$ fb for $m_h\simeq 100$ GeV instead of the
naive estimation of $\simeq 300$ fb obtained by scaling the single-top
cross section by $\sigma(t\bar t h)/\sigma(t \bar t)$. The main reason
for this is a particularly strong destructive interference between the
two dominant amplitudes, associated with the diagrams shown in
Fig.~\ref{fig:th_sm}. It can be shown that each diagram contains a
term proportional to $m_t$ which violates unitarity at high energies
and cancels in their sum.

\FIGURE[t]{
\epsfig{file=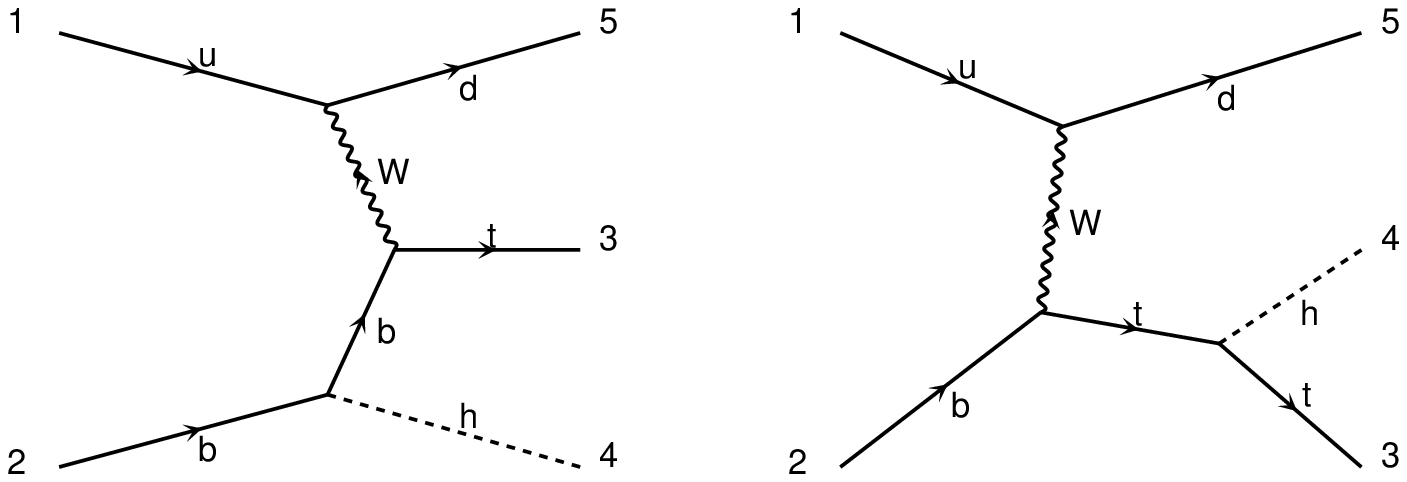, width=8.5cm}
\epsfig{file=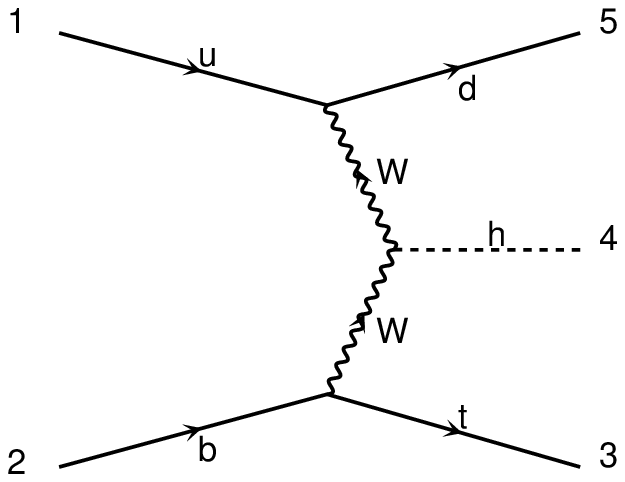, width=4cm}
\caption{The diagrams contributing to single-top and Higgs associated
production. The first diagram is negligible in the SM and in type I
2HDMs while the second one is negligible in the MSSM and in type II
2HDMs.}
\label{fig:th_sm}
}

\FIGURE[t]{
\epsfig{file=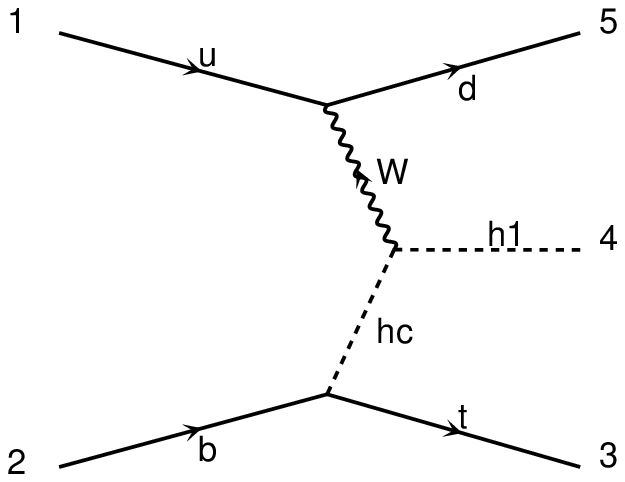, width=4cm}
\caption{Extra diagram contributing to single-top associated Higgs production
in the MSSM and 2HDM.}
\label{fig:th_2hdm}
}

\begin{figure}
\begin{center}
\begin{tabular}{p{7cm}r}
\includegraphics[width=0.5\textwidth]{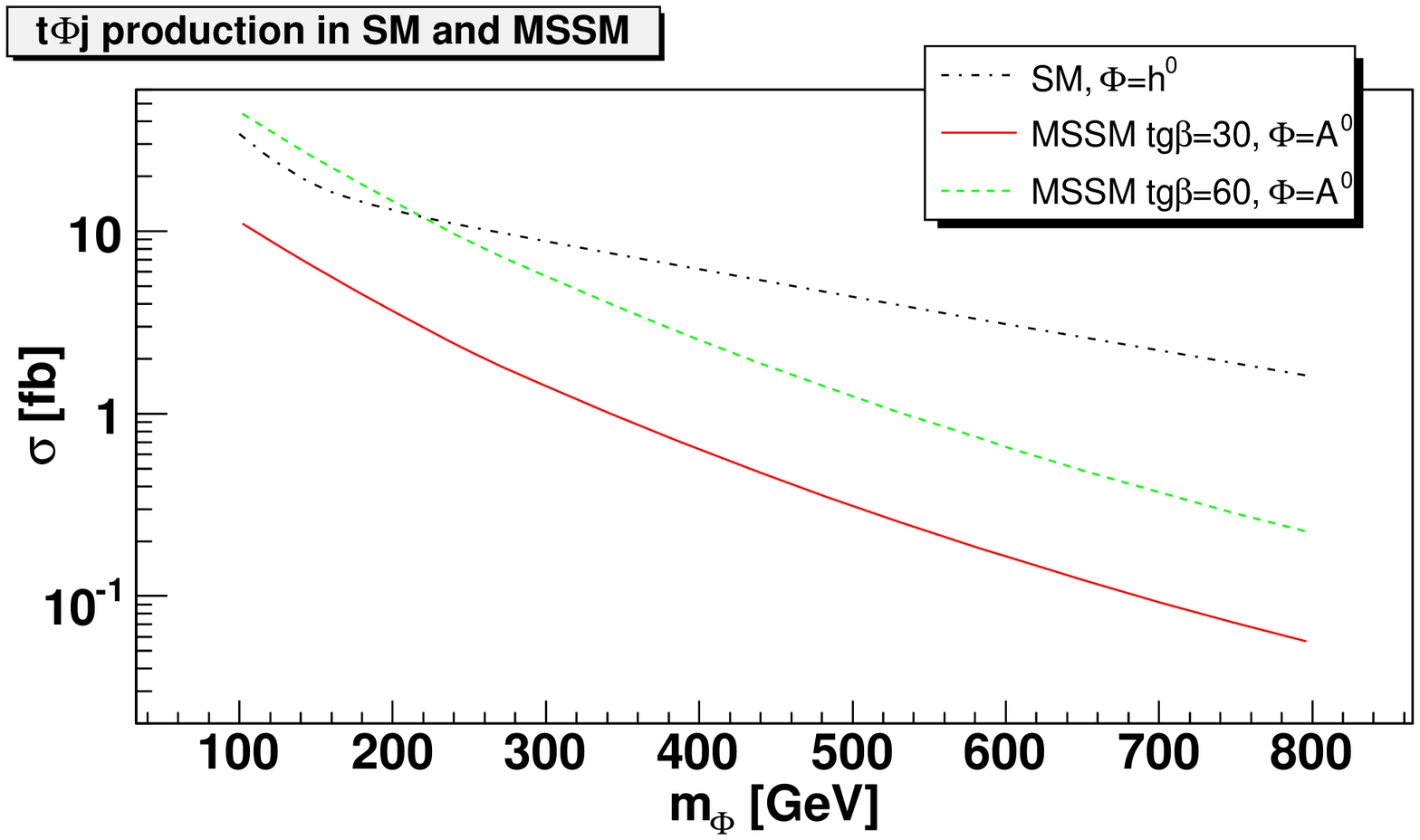}
&
\includegraphics[width=0.5\textwidth]{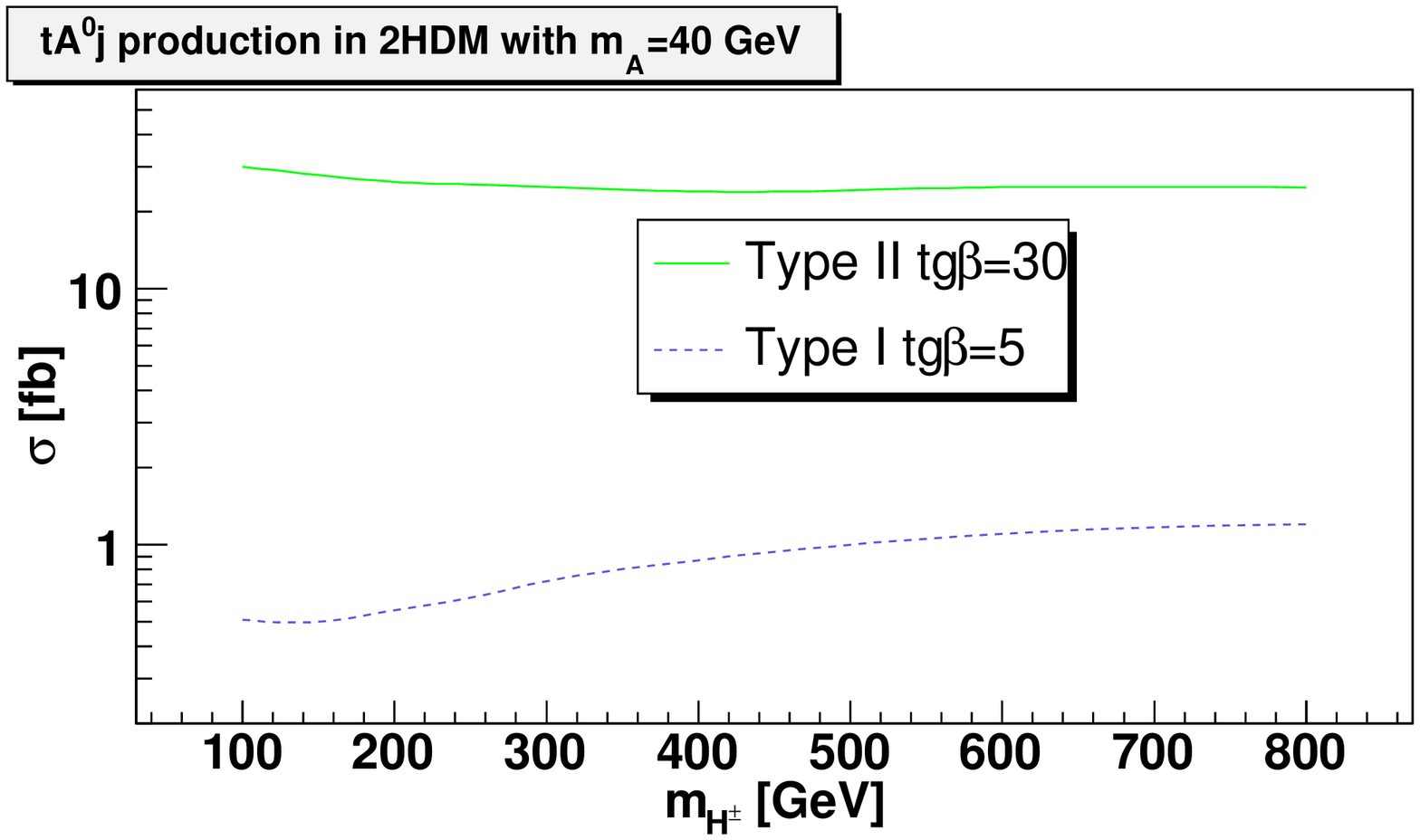}
\end{tabular}
\caption{Cross sections of single top associated production of
different Higgs bosons. On the left, the SM Higgs boson production
cross section is shown together with the MSSM pseudoscalar $A^0$
production cross section as a function of their masses for two
different $\tan\beta$ values. On the right, the 2HDM pseudoscalar
$A^0$ production cross section is shown as a function of $m_{H^\pm}$
for type I and type II Yukawa couplings. A minimal $p_T$ of 20 GeV and
a maximal rapidity of 2.5 is assumed for the jets. The factorisation
and renormalisation scales are both set equal to $m_\phi$. The $b$
quark mass involved in the Yukawa coupling as well as its pole mass
(in order to deal consistently with unitarity cancellations) are equal
to the running mass at $m_\phi$. The PDF used is CTEQ6L1.}
\label{fig:crossec}
\end{center}
\end{figure}

So, even though the sign of this interference can be guessed from
unitarity requirements, its absolute magnitude at moderate energies,
compared to the masses involved (\eg, the top mass), is
surprising. The integrated squared amplitude for each of these diagram
seperately is indeed up to three times larger than the total cross
section. The total SM cross section at LHC as a function of $m_h$ is
shown in Fig. \ref{fig:crossec}.

Such a strong cancellation also takes place in the
MSSM~\cite{Maltoni:2001hu}, but in this case more aspects have to be
taken into account. To illustrate this, let us consider the decoupling
regime where the mixing between $h^0$ and $H^0$ is small. For the
light Higgs boson, the situation is similar to the SM. The $WWh^0$
coupling is close to the SM value and in the limit of large
$\tan\beta$ where $h^0$ couples mainly to $b$, the $b\overline{b}h^0$
coupling is close to the $t\overline{t}h$ of the SM. The amplitude of
the diagram where the Higgs boson is emitted from the initial state
$b$ quark appears to be slightly suppressed compare to the one
associated to the diagram where it comes from the final state $t$
quark (up to a factor $2$ for identical Yukawa couplings), but the
overall impact on the physical cross section stays small, or is even
positive, since the negative interference previously described in the
SM is also decreased by the same factor. For the heavier scalar $H^0$
the coupling $WWH^0$ almost vanishes and one could then expect an
enhancement of the total cross section.  However there is an
additional diagram involving a charged Higgs boson (see
Fig.~\ref{fig:th_2hdm}) which has to be taken into account due to the
large $W^\pm H^\mp H^0$ coupling. This diagram leads to a amplitude of
roughly the same order and the same sign as the one involving only the
$W$ boson and thus no particular enhancement is observed. The
situation is similar for the pseudoscalar $A^0$ (see
Fig.~\ref{fig:crossec}) even though in this case the $WWA^0$ coupling
is strictly zero due to $CP$ invariance.

Another possibility to get an increase of the cross section for the
latter process
would be to consider a 2HDM where the pseudoscalar $A^0$ is relatively
light and where the charged Higgs pair is much heavier (\textit{e.g.},
see \cite{Gerard:2007kn}) so that the negative interference cannot
occur (\ie, the amplitudes associated with the SM like diagrams in
Fig.~\ref{fig:th_sm} are dominant). This, of course, cannot occur in
the MSSM where the masses of $H^\pm$ and $A ^0$ are linked at tree
level through the mass relation
$m_{H^\pm}^2=m_{A^0}^2+m_{W^\pm}^2$. The resulting cross section for
$m_{A^0}=40$ GeV is plotted in Fig.~\ref{fig:crossec} both in case of
type I and of type II 2HDM as a function of the charged Higgs mass. An
enhancement at high $m_{H^\pm}$ is observed in the 2HDM type I case
but the overall cross section stays much smaller than the SM one due
to the reduced top quark Yukawa coupling. In type II models, the
effect of varying $m_{H^\pm}$ is quite small. At low $m_{H^\pm}$, the
diagram involving the charged Higgs boson should contribute but its
squared amplitude is more or less of the same order of magnitude as
the negative interference it creates, so that its total contribution
is negligible. Like in the MSSM, for all values of $m_{H^\pm}$, the
overall cross section appears to be slightly suppressed, more or less
by a factor $2$, compared to the SM when the pseudoscalar is emitted
from the initial state $b$ quark.

To conclude, the cancellations in single top associated Higgs
production expected from the unitarity of the models considered
together with the suppressions due to different emission
configurations are very effective in reducing the cross section to a
value close the SM one, which is probably too small to be successfuly
measured at LHC.

\subsection{Comparison of strong SUSY pair production at the SPS points}

As a simple example showing some of the power of the SUSY
implementation in MadGraph/MadEvent~4, we have chosen to compare
results from inclusive strong SUSY pair production for the ten
Snowmass (or SPS) benchmark parameter points~\cite{Allanach:2002nj}. In
particular, the example makes use of the following new features:

\begin{itemize}
\item Implementation of the SUSY particle and
interaction content \cite{Cho:2006sx} and the possibility to read SLHA
files with mass, coupling and mixing information.
\item Inclusion of more than one process in the MadGraph
generation. The different processes included in this way are then
produced in proportion to their cross section, so the event file
produced automatically gets the correct mix of unweighted events.
\item The possibility to define multi-particle labels also for scalar
particles.
\item The interfaces to Pythia for decay and hadronization, PGS
for detector simulation and ROOT for event analysis.
\end{itemize}

\subsubsection{Setup and generation}

To easily be able to see the relative importance of different groups
of subprocesses, we generate the process using the {\tt proc\_card.dat}
shown in Fig.~\ref{fig:susy-proc-card}. In this card we differentiate
between gluino pair production, squark of the first and second families
pair production, third familiy squark pair production, associate
production of gluinos and squarks, and associate prodution of third
family squarks with first and second family squarks. In the
generation all these subprocesses (in total 497 different
subprocesses, since we make a distinction among the different
squark flavors) are automatically generated in the correct
proportions according to their relative cross sections.

\begin{figure}
\footnotesize{
\begin{verbatim}
# Begin PROCESS # This is TAG. Do not modify this line
pp > gogo   @1 #  Process 
QCD=2          # max qcd order
QED=0          # max qed order
end_coup       # the coupling list is over

pp > S1S1   @2 #  Process 
QCD=2          # max qcd order
QED=0          # max qed order
end_coup       # the coupling list is over

pp > S2S2   @3 #  Process 
QCD=2          # max qcd order
QED=0          # max qed order
end_coup       # the coupling list is over

pp > S1go   @4 #  Process 
QCD=2          # max qcd order
QED=0          # max qed order
end_coup       # the coupling list is over

pp > S2go   @5 #  Process 
QCD=2          # max qcd order
QED=0          # max qed order
end_coup       # the coupling list is over

pp > S1S2   @6 #  Process 
QCD=2          # max qcd order
QED=0          # max qed order
end_coup       # the coupling list is over

done           # the process list is over
# End PROCESS  # This is TAG. Do not modify this line
# Begin MODEL  # This is TAG. Do not modify this line
mssm 
# End   MODEL  # This is TAG. Do not modify this line
# Begin MULTIPARTICLES # This is TAG. Do not modify this line
p  uu~dd~ss~cc~bb~g 
S1 ulul~urur~dldl~drdr~slsl~srsr~clcl~crcr~
S2 b1b1~b2b2~t1t1~t2t2~
# End  MULTIPARTICLES # This is TAG. Do not modify this line
\end{verbatim}}
\caption[]{\label{fig:susy-proc-card}The proc\_card.dat used for the 
generation of all strong $2\to2$ SUSY processes. The processes are
grouped such that gluino pair production is in the first group, first
and second generation squark pair production is in the second group,
third generation in the third and associated production in the last
three groups. Please note that further subdivisions have been made in
Table~\ref{tab:SPS-xsecs}.}
\end{figure}

We next generate events at the LHC for each of the ten SPS points, and
run the events through Pythia 6.409 \cite{Sjostrand:2006za}, which
decays the SUSY particles and performs parton showering and
hadronization, and finally through PGS 4 \cite{PGS4}, a general-purpose
detector simulation. We used the default LHC setup for PGS 4. The
parameter cards for the points were generated using SoftSusy 2.0.5
\cite{Allanach:2001kg} and
Sdecay 1.1a \cite{Muhlleitner:2003vg} and are available from the MG/ME 4
web sites. The results of the generation are:

\begin{enumerate}
\item The total cross section for strong SUSY pair production.
\item Cross sections for all the individual subprocesses, and summed
cross sections for the different groups of subprocesses.
\item Files with unweighted events at parton level, decayed and
hadronized level and detector reconstruction level.
\item ROOT files for event analysis at all three levels.
\end{enumerate}

\subsubsection{Comparison of the results for the SPS points}

The cross sections and relative importance of the contributing
subprocesses for the different SPS points are presented in Table 4.
The relative cross sections for the different subprocesses are
calculated by MadEvent at leading order, but we should be able to take
the results as a good indication of which processes are most important
to take into account at each point.

\begin{table}\begin{center}
\begin{tabular}{|c|c|c|c|c|c|c|c|c|c|}
\hline
SPS pt. & $\sigma$ (pb)
&       $\sigma_\mathrm{NLO}$ (pb)
&       $\tilde g \tilde q$
&       $\tilde q\tilde q$
&       $\tilde g\tilde g$
&       $\tilde t\tilde t^*$
&       $\tilde b\tilde b^*$
&       $\tilde g\;\mathrm{sq}^{(*)}$
&       $\mathrm{sq}^{(*)}\mathrm{sq}^{(*)}$\\
\hline
1a & 42.6 & 52.7 & 45 & 11 & 12 & 3.4 & 1.4 & 9.7 & 17\\
\hline
1b & 3.39 & 4.51 & 44 & 21 & 6.6 & 2.4 & 1.3 & 6.4 & 18.5\\
\hline
2 & 1.48 & 2.70 & 29 & 3.1 & 56+7.0 & 0.55 & 0.073 & 3.1 & 0.8\\
\hline
3 & 3.81 & 4.87 & 44 & 21 & 6.9 & 2.5 & 1.1 & 6.1 & 18\\
\hline
4 & 10.0 & 13.9 & 47 & 14 & 13 & 2.6 & 1.5 & 8.3 & 17\\
\hline
5 & 35.5 & 39.2 & 29 & 9.2 & 6.1 & 33+4.7 & 1.1 & 5.4 & 11\\
\hline
6 & 14.5 & 20.5 & 45 & 15 & 9.7 & 2.3 & 1.2 & 8.4 & 17\\
\hline
7 & 3.08 & 4.00 & 45 & 22 & 7.5 & 1.0 & 0.7 & 6.2 & 17\\
\hline
8 & 2.34 & 3.52 & 50 & 13 & 23 & 0.4 & 0.3 & 6.4 & 6.7\\
\hline
9 & 0.425 & 0.506\ & 42 & 31 & 5.4 & 2.5 & 1.0 & 3.9 & 14\\
\hline
\end{tabular}\end{center}
\caption{\label{tab:SPS-xsecs}Cross sections for the ten SPS points and
the contributions, in \%, from different subprocesses for each point. In the
process definitions, $\tilde q=\tilde u_{L,R},\tilde d_{L,R}$. For
$\tilde g\tilde g$ and $\tilde t\tilde t^*$ production, the dominant
channel is gluon fusion, with $q\bar q$ annihilation contributing to less than
10\% of the numbers. The exceptions are point 2 and 5, respectively,
where we have separated the $gg$ and $q\bar q$ channels by the ``+''
sign. The last two columns show the contributions from associated
gluino-(anti)squark production where the processes in column 4 are
excluded, and (anti)squark-(anti)squark production where the processes
in column 5, 7 and 8 are excluded. The NLO cross sections are
calculated using Prospino 2
\cite{Beenakker:1996ch,Beenakker:1997ut,prospino2}.}
\end{table}

A striking feature of Table~\ref{tab:SPS-xsecs} is that between 68\%
(SPS1a) and 92\% (SPS2) of the total cross section is due to only the
production of valence quark partners, gluinos, and in some cases top
partners, constituting $\sim 25$ processes out of almost 500. This is
of course due to the dominance of the valence $u$ and $d$ quarks over
other quark species in the parton distributions of the proton, which
is larger the heavier the squarks are.

\begin{figure}
\begin{center}
\includegraphics*[width=10cm]{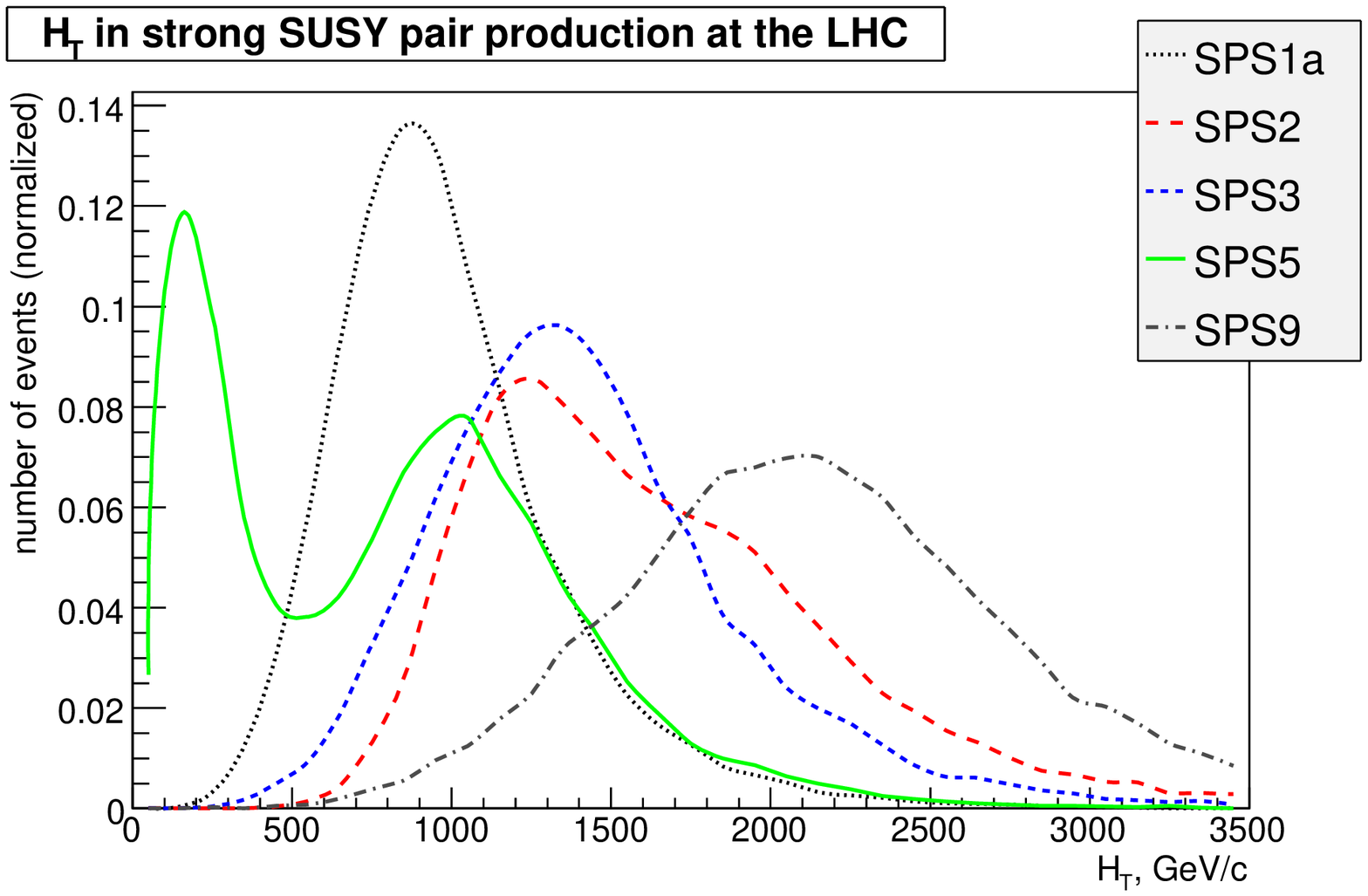}
\end{center}
\caption{\label{fig:sps-ht}$H_T$ distribution for SPS
points 1a, 2, 3, 5 and 9. The peak or peaks of the distribution is
correlated with the masses of the produced SUSY particles.}
\end{figure}

We also look at kinematical event distributions for the different
points, after decay, hadronization and detector simulation. One of the
most suggestive distributions is the total transverse event energy, $H_T$,
defined as $H_T=\sum p_\perp^\mathrm{jets}+
\sum p_\perp^\mathrm{leptons}+ \not \!\! E_\perp$, since it gives an
indication of the mass scale of the produced particles. In
Fig.~\ref{fig:sps-ht} the $H_T$ distributions are shown for a reduced
set of SPS points. We can clearly see how the position of the peak (or
peaks) is correlated with the masses of the particles produced, see
Table~\ref{tab:mass-peak-rel}. For the ten SPS points, the peak
positions are at about 65-85\% of the sum of the masses of the produced
particles, with a slightly increasing ratio as the masses
increase. The exception is the very low mass stop pair production peak
at SPS5, where the peak position is below 40\% of the sum of the produced stop
masses. This is due to the large proportion of the event energy
carried away by the two LSPs, which largely balance in the detector so
that the total missing energy detected is relatively small. If we add
$2 m_{\chi_1^0}$ to the peak positions, they reach between 
85\% and 100\% of the mass of the produced particles, again with a
slightly rising trend going to higher masses.

\begin{table}
\begin{center}
\begin{tabular}{|l|c|c|c|c|c|}
\hline
SPS point & 1a & 2 & 3 & 5 & 9 \\
\hline
Peak position(s)(GeV) & 800 & 1200 / 1900 & 1300 & 200 / 1100 & 2200 \\
\hline
Particle masses (GeV) & 
        \parbox{15mm}{$\tilde g$: 600 \\ $\tilde q$: 550 \\
        $\chi_1^0$: 100} &
        \parbox{15mm}{$\tilde g$: 780 \\ $\tilde q$: 1550 \\
        $\chi_1^0$: 120} &
        \parbox{15mm}{$\tilde g$: 930 \\ $\tilde q$: 830 \\
        $\chi_1^0$: 160} &
        \parbox{15mm}{$\tilde t_1$: 260 \\ $\tilde g$: 720 \\ 
        $\tilde q$: 650 \\ $\chi_1^0$: 120} &
        \parbox{15mm}{$\tilde g$: 1290 \\ $\tilde q$: 1260 \\
        $\chi_1^0$: 200} \\
\hline
\end{tabular}
\end{center}
\caption{\label{tab:mass-peak-rel} The positions of the peaks in the
$H_T$ distributions and the masses of the particles mainly produced, as
well as the LSP, for the SPS points 1a, 2, 3, 5 and 9.}
\end{table}

In the present study, we did not make use of the most important feature
of MadGraph/MadEvent: its efficiency in calculating multiparticle
final states. In particular, we used Pythia to perform the decay of
the SUSY particles produced. For the scalar quarks this should be a
reasonable approximation, since no spin correlations are expected
(although there might still be effects from interference or
Breit-Wigner curve distortions in some parameter regions). However, if
we want to study angular distributions where spin correlations might
be important, the gluino decay should be done in MadEvent. For this
kind of refinements, efficiency will start to become an important
factor. It is therefore better to use different treatments for the
different parameter points, in order to make an optimized choice for
which processes to include at each point. So this should be seen
rather as a preparatory study for such a more elaborate analysis.

\subsection{Inclusive $W+$jets matched samples: comparison with the Tevatron data}
\label{sec:wjets}

\FIGURE[t]{
\includegraphics*[width=13cm]{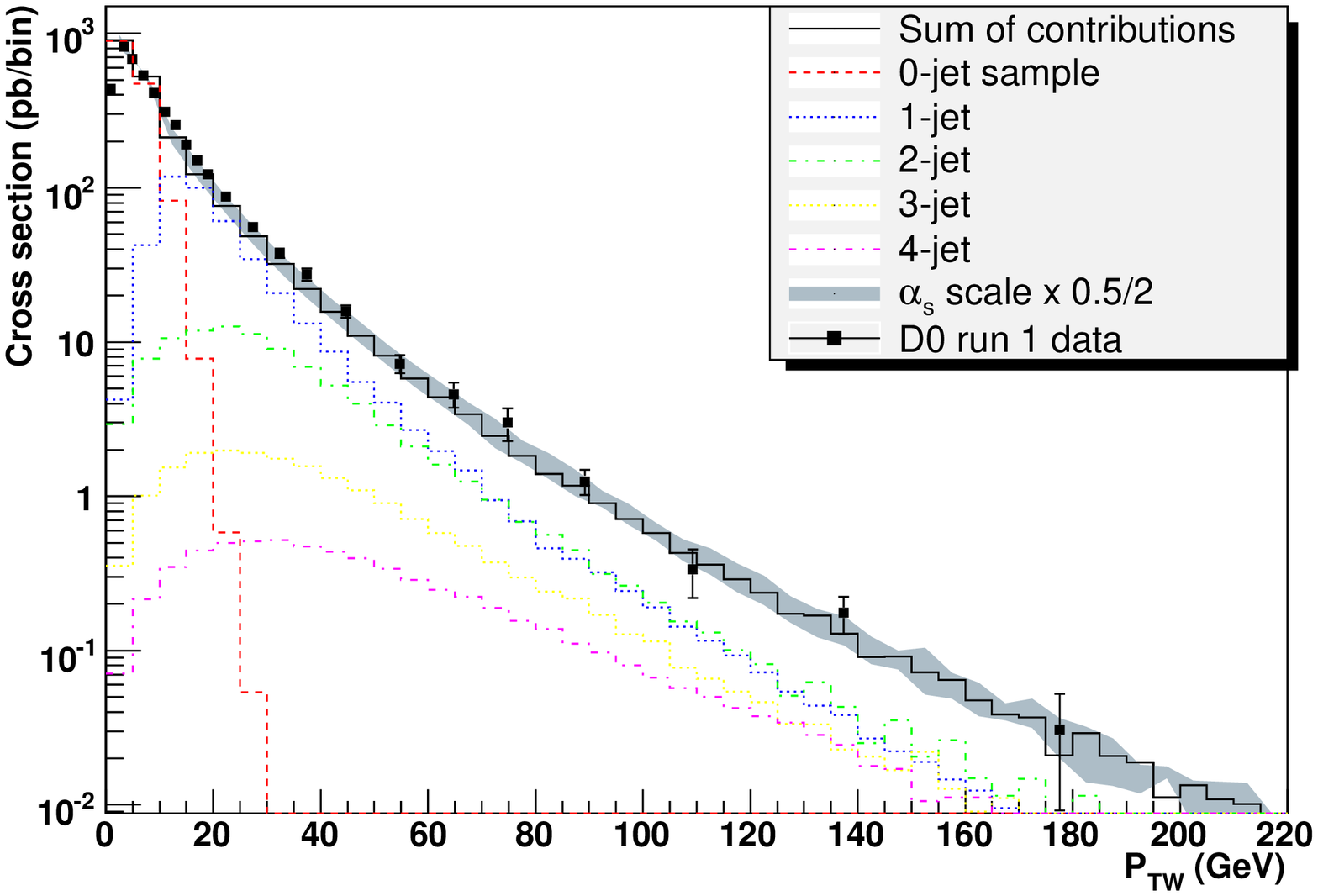}
\includegraphics*[width=13cm]{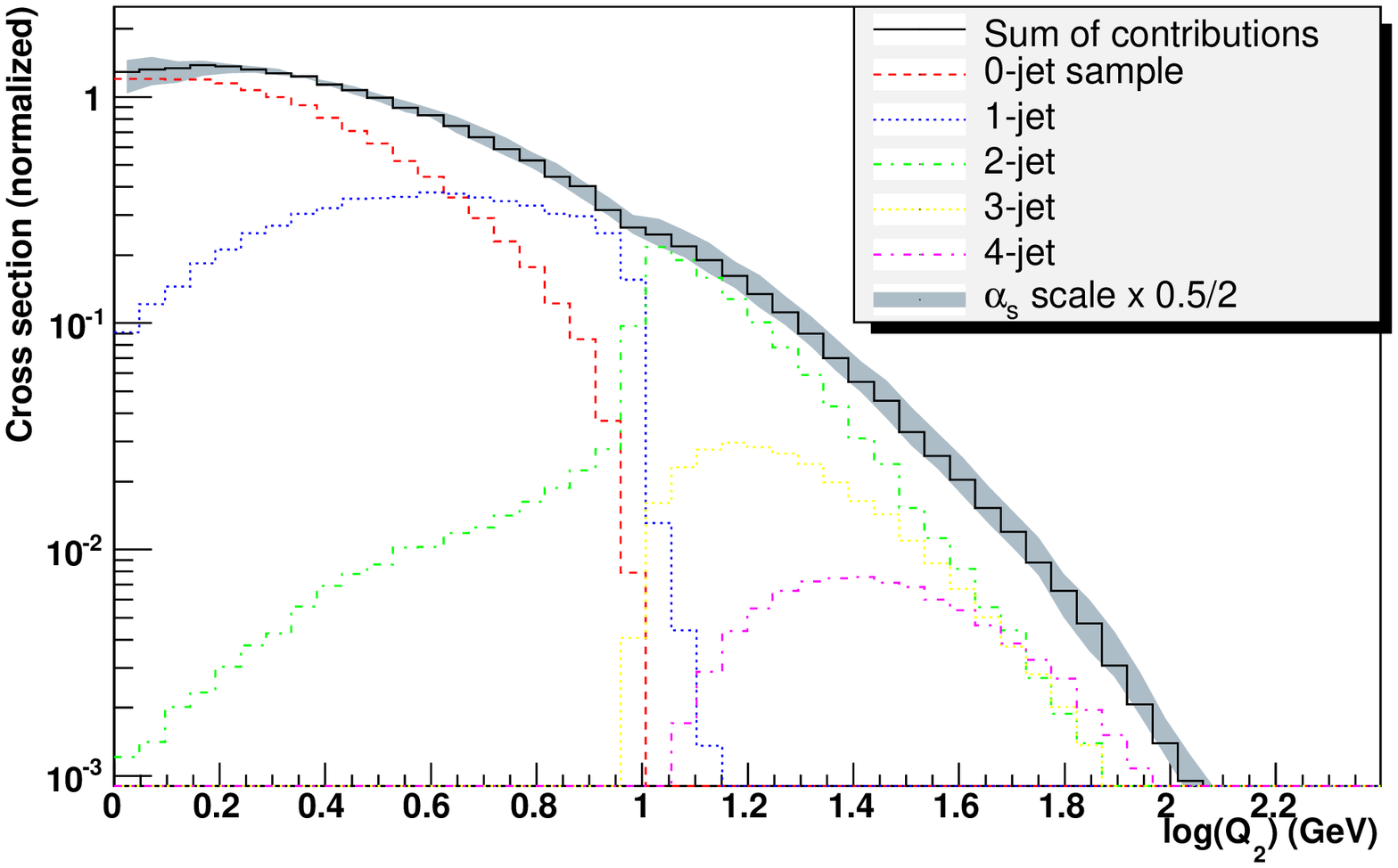}
\caption{\label{fig:wjets} The $p_\perp$ of the $W$ boson and the
parton-level differential $1\to2$ jet rate for $W$+jets production at
the Tevatron, using matching of matrix elements and parton showers
with MadEvent and Pythia. Please refer to the text for further
discussion of the plots.}
}

At the Tevatron, as well as the LHC, many interesting signals include
an isolated lepton, missing energy and hard jets. This means that weak
boson production plus QCD jets is an important background. Until
recently, this background was estimated using parton shower Monte
Carlo's such as Pythia, where the largest jet multiplicity is $W^\pm+1$
jet, and additional jets are produced by parton showering. As
discussed in Section \ref{sec:matching}, this gives a well-founded
description of jets with small transverse momentum, and jets close in
phase space, but for well separated hard jets it misses important
non-logarithmic effects such as interference between diagrams. For
that case, matrix element calculations are necessary. In order to
simultaneously describe well-separated jets of different
multiplicities, some kind of matching is needed between these
descriptions.

$W^\pm$ and $Z$ boson production with multiple jets is for several
reasons an excellent testing ground for matching procedures; there is
a simple ``central'' $2\to1$ process, there is a hard scale set by the
mass of the vector boson, and there is data from the Tevatron to
compare to. A comparison between several implementations of matrix
element--parton shower matching of $W+\mathrm{jets}$ production at
hadron colliders is in progress \cite{matchcomp}.

In order to assess the robustness and flexibility of the method, there
are several possibilities for variations. Keeping to the principal
matching method used in MadEvent and Pythia, a modified MLM method
using the $k_\perp$-clustered jet definition, the most prominent
variable is the so-called matching scale, used as cutoff between the
matrix element and parton shower descriptions for jet
production. Other natural variables are the scales for the running of
$\alpha_s$, which can be varied in the parton shower and/or in the matrix
element generation, and the scale used in the parton
densities, which might also be varied independently. In
Fig.~\ref{fig:wjets}, we have chosen a fixed matching scale of 10 GeV
and varied the scale for $\alpha_s$ in the matrix element generation and the partons showers
by a factor 2 up and down from the default scale, set by the $k_\perp$
in the clustering corresponding to that power of $\alpha_s$.

Fig.~\ref{fig:wjets}a shows the $p_\perp$ of the $W$ boson in $W$
production at the Tevatron, including hadronization of quarks and
gluons but without underlying event simulation. The full black line
shows the matched sum of the different jet multiplicity contributions,
which are shown in different colors and line styles. For very low
$p_{\perp W}$ the pure parton shower dominates, but from the matching
scale and up, the different jet multiplicities become more and more
important, until around 150 GeV, where all the four included
multiplicities are of similar order of magnitude. The band in the plot
corresponds to adding a factor $\frac12-2$ in the arguments of
$\alpha_s$ in the matrix element generation only. It can be seen that
the simulation with default parameters reproduces the shape of the
Tevatron data \cite{Abbott:2000xv} (the normalization has been fitted
by 20\%).

Fig.~\ref{fig:wjets}b shows the differential jet rate going from
$1\to2$ jets, after parton showers but before hadronization. It is
included as an example, showing very clearly the transition between
parton showers (in this case from the 0-jet and 1-jet samples), below
the matching scale $\log Q_2 = \log k_\perp^\mathrm{cut} = 1$, and
matrix elements (from the 2-jet sample and upward) above the matching
scale, as well as the relative smoothness of the transition. The jet
rate is the variable used to perform the matching, although here only
particles with $|\eta|<2.5$ are included in the jet definitions,
explaining the low-end tails of the higher multiplicity samples.

\section{Conclusions and Outook}
In this paper we have presented the new release of the
MadGraph/MadEvent package which aims at providing a tool to perform
the simulation of both signal (Standard Model and beyond) and
backgrounds within one framework. It is worth summarizing the key
points of our approach.

First, MG/ME is a user-driven Monte Carlo. Our package {\it does not
contain a library of processes} but creates the code specific to the
user's requests. It {\it contains an easy-to-extend library of physics
models} which the user can choose from, including the Standard Model,
the Higgs effective-theory, the MSSM, the most general 2HDM.

Second, event generation with MadGraph is available from the web, in
addition to the downloadable source code.  All
functionalities of our package can be accessed directly via a web
interface without the need of any installation/compilation. User's
code creation, event generation and detector simulation can be
controlled by filling (or uploading) simple input cards and the
results in the form of event files and plots can be downloaded from
the user's personal database.  MG/ME event generation is based on an
algorithm which parcels the overall work in many small independent
jobs. At present, our resources include three medium-sized PC clusters
that are available and open to the public for code creation and the
generation of (limited-size) event samples.  The web server software
is portable and can be easily installed on any cluster running PBS or
Condor. Extension to work with the Grid software are under
development. The source code for this, and support for
experimental as well as theoretical groups is available upon request
to the authors.

MG/ME is designed to be a tool to ease the communication between
theorists and experimentalists. Any theorist, model builder or
phenomenologist, can implement and carefully test his/her own field
theoretic model in MG/ME. The corresponding phenomenology at
colliders, such as the LHC or an ILC, can be easily studied up to the
simulated detector level by themselves and/or with the help of
experimental colleagues.  Both signal and background can be simulated
within the same framework. Eventually, the experimental groups can
pick it up and promote it to a ``exp-grade" analysis by using their
detector specific tools.

The advantages of having ``one framework for all" is manifest for both
theorists (model builders) and experimentalists. For the former, it is
important that the tool is very flexible, the learning curve is mild
and no specific programming skills are needed.  In particular, all the
tedious and error-prone (and sometimes even difficult) tasks like
matrix element creation, cross section integration and event
generation, are automatically taken care of.  On the experimental
side, the insertion of a code in the simulation chain is normally a
time-expensive process which involves testing and validation.  For
MG/ME this can be done, once for all, through a small set of simple
Standard Model processes where comparison with available MC can be
quickly done.

The further developments that are planned for the package are also
towards building a framework suitable for theorists/experimentalists
interactions.

In the current version, we have added a semi-automatic framework to
implement new physics models. Even though this framework is quite
general and very simple to use, it has two main drawbacks. First, the
Feynman rules of a given model have to be available. This is trivial
for very simple models but it can be become rather tedious and
cumbersome for richer or more complete models (such as SUSY). Second,
the range of models that can be easily implemented is still limited by
the kinematic form of the interactions in HELAS, which has to be
similar to those present in the Standard Model. We are currently
working on a package that starting from a generic Lagrangian generates
the Feynman diagrams in a MadGraph compatible format and on a more
flexible version of HELAS.

One feature that is under testing and will be soon included is the
possibility of selecting the so-called decay chains in the generation
of very rich multi-particle final states arising from the multi-staged
decays of heavy particles in arbitrary models. By using a
matrix element-based approach to decays, all information about
possible correlations, such as those coming from the spin of the
intermediate resonances, is available and can be exploited in the
analysis.

Another line of development is towards building general analysis tools
that can maximally exploit the information encoded in the matrix
element to improve the accuracy or the sensitivity of measurements of
key parameters (such as masses or couplings). These techniques have
been successfully employed in several LEP and the Tevatron analyses,
but always on a process specific basis. A general and automatic
approach would be certainly welcome.

\section*{Acknowledgements}
We are thankful to the many people that during these years have
directly or indirectly helped and/or supported the developement
MadGraph/MadEvent. In particular, Michelangelo Mangano and the Alpgen
boys, Frank Krauss and the SHERPA kids, the Pythia and Herwig teams
and in particular Torb\"orn Sj\"ostrand, Steve Mrenna. We are in debt
with our golden collaborator Kaoru Hagiwara for his continuous contributions to the project, from
SMadGraph to the HELAS development. On the experimental side, we thank
Henry Frisch for being always among our most enthusiastic (and patient)
users, Bruce Knuteson for being our most skilled hacker, Tony Liss,
Sasha Nikitenko, and many many others among D0, CDF, ATLAS, and CMS
physicists. We would also like to thank all the members of CP3 for
the great atmosphere and enviroment that foster our efforts. Finally,
we are greateful to David Lesny, Larry Nelson (UIUC), 
Fabrice Charlier, Thomas Keugten (UCL) and   
Roberto Ammendola, Filippo Palombi, Nazario Tantalo (Centro Fermi) for 
their restless and reliable technical support.
Our human and computing resources are supported by 
the US National Science Foundation (Contract number NSF PHY 04-26272 ), 
Centro Fermi - Rome, Universit\'{e} Catholique de Louvain, 
Institut Interuniversitaire des  Sciences Nucl\'eaires, Belgian Interuniversity Attraction Pole P6/11 
and the Swedish Research Council.      

\newpage
\bibliography{database}
\end{document}